\documentclass[12pt]{article}
\usepackage{subcaption}
\usepackage[export]{adjustbox}
\usepackage{bm}
\usepackage[figuresright]{rotating}
\usepackage{color}
\usepackage{mathrsfs}
\usepackage{booktabs}
\usepackage{enumerate}
\usepackage{array}
\usepackage{float}
\usepackage{epstopdf}
\usepackage{mathrsfs}
\usepackage{amssymb}
\usepackage{amsthm}
\usepackage{amsmath}
\usepackage{algorithm}
\usepackage{algorithmic}
\usepackage[nodisplayskipstretch]{setspace}
\usepackage{natbib}
\usepackage{multirow}
\usepackage{url}

\makeatletter
\def\singlespace{\def\baselinestretch{1}\@normalsize}

\newtheorem{theorem}{Theorem}
\newtheorem{remark}{Remark}
\newtheorem{definition}{Definition}

\newtheorem{proposition}{Proposition}
\newtheorem{corollary}{Corollary}

\begin{document}

\def\spacingset#1{\renewcommand{\baselinestretch}%
{#1}\small\normalsize} \spacingset{1}

\title{\bf Nonparametric Statistical Inference via Metric Distribution Function in Metric Spaces}
\author{Xueqin Wang\thanks{The author is partially supported by National Natural Science Foundation of China grants No.12231017, the National Key R\&D Program of China No.2022YFA1003803, and National Natural Science Foundation of China grants No.72171216, 71921001, and 71991474.}, Jin Zhu, Wenliang Pan\thanks{Pan's research is partially supported by the National Natural Science Foundation of China Grants No.12322113 and No.12071494.}, Junhao Zhu and  Heping Zhang\thanks{Zhang's research is supported in part by NIH grants HG010171 and MH116527 and NSF grant DMS 2112711. } \\
{University of Science and Technology of China, Sun Yat-Sen University}\\{London School of Economics and Political Science, Chinese Academy of Sciences}\\ {Yale University}\vspace{-14pt}\thanks{Data used in the preparation of this article were obtained from the Alzheimer's Disease Neuroimaging Initiative database (\protect\url{adni.loni.usc.edu}). As such, the investigators within the ADNI contributed to the design and implementation of ADNI and/or provided data but did not participate in the analysis or writing of this report. A complete listing of ADNI investigators can be found at: \protect\url{adni.loni.usc.edu/wp-content/uploads/how_to_apply/ADNI_Acknowledgement_List.pdf}. We are grateful to Prof. Hongtu Zhu for generously sharing the preprocessed ADNI dataset with us. All authors contributed equally to this work. }}


\date{}
\maketitle

\bigskip


\begin{abstract}
The distribution function is essential in statistical inference and connected with samples to form a directed closed loop by the correspondence theorem in measure theory and the Glivenko-Cantelli and Donsker properties. This connection creates a paradigm for statistical inference. However, existing distribution functions are defined in Euclidean spaces and are no longer convenient to use in rapidly evolving data objects of complex nature. It is imperative to develop the concept of the distribution function in a more general space to meet emerging needs. Note that the linearity allows us to use hypercubes to define the distribution function in a Euclidean space. Still, without the linearity in a metric space, we must work with the metric to investigate the probability measure. We introduce a class of metric distribution functions through the metric only.
We overcome this challenging step by proving the correspondence theorem and the Glivenko-Cantelli theorem for metric distribution functions in metric spaces, laying the foundation for conducting rational statistical inference for metric space-valued data. Then, we develop a homogeneity test and a mutual independence test for non-Euclidean random objects and present comprehensive empirical evidence to support the performance of our proposed methods.
\end{abstract}
\bigskip


\newpage
\spacingset{1.9}


\section{Introduction}\label{sec:introduction}
Nowadays, many statistical applications study non-Euclidean data.
Typical data examples include symmetric positive definite (SPD) matrices \citep{smith2013functional}, the Grassmann manifold \citep{HongMarc2016},
the shape representation of corpus callosum \citep{cornea2017regression},
samples of probability density functions in Wasserstein spaces \citep{petersen2021wasserstein},
paleomagnetic directional data in hyperspheres \citep{wood2019scaled}.

Analysis of non-Euclidean objects is challenging \citep{petersen2021wasserstein, cornea2017regression, wood2019scaled}.
A common strategy is to embed non-Euclidean data objects into a Hilbert or more general metric space before the analysis.
When the non-Euclidean data objects can be embedded in a metric space but not a Euclidean space, metric (distance)--based methods can be applied.
Many methods exist or are being developed in statistics and machine learning, e.g.,
\cite{szekely2004testing,szekely2007measuring,bjorn2019distancemultivariance, pan2018ball, muller2019frechet, daiTukeyDepthObject2021}.
Assessing the uncertainty following the use of the existing methods to analyze non-Euclidean data is important but difficult due to the absence of a fundamental concept in metric spaces analogous to the distribution function (DF) in Euclidean spaces.

\begin{figure}[ht]
	\vspace*{-5pt}
	\begin{center}
		\includegraphics[scale=1.0]{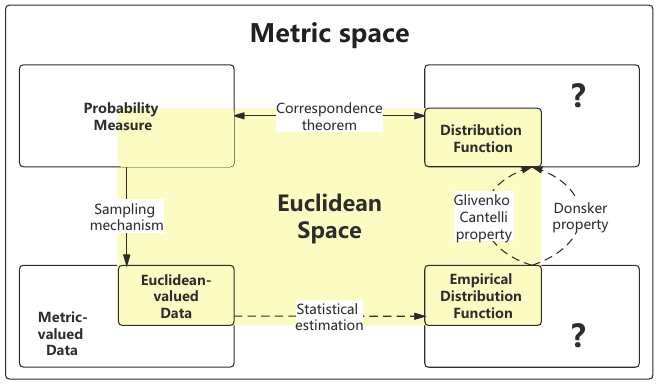}
	\end{center}
	\vspace*{-10pt}
	\caption{Conceptual diagram of statistical inference paradigm. The solid arrows indicate a conceptual deduction, and the dashed ones are a statistical approximation.}
	\label{inference-framework}
	\vspace*{-15pt}
\end{figure}

DF relates theory to the real world in statistical inference, allowing us to conclude the data \citep{10.1214/aos/1176344552}.
The DF is defined to uniquely determine the Borel probability measure of a random vector (or a scalar) according to the correspondence theorem \citep{halmos1956measure}.
Given observed data, the DF can be well estimated by the empirical distribution function (EDF).
As illustrated in Figure~\ref{inference-framework}, the DF and observed sampled data are linked to form a directed closed loop by the correspondence theorem in measure theory and the Glivenko-Cantelli theorem in statistical inference. This connection creates a paradigm for statistical inference.

The properties and applications of EDF have been systematically investigated as a prominent field in mathematical statistics for a century, and 
many statistical methods are, in fact, functional of EDF. Examples include the Cram{\'e}r-von Mises test \citep{darling1957the} for the equality of two unknown DFs and Hoeffding's independence test \citep{hoeffding1948independence}  for two random data samples. Thus, it is reasonable to anticipate the importance of generalizing the 
concepts of DF and EDF to metric spaces to have a basic foundation for the methods we may use to analyze non-Euclidean data objects.

In this paper, we introduce a quasi-DF to serve as the cornerstone of nonparametric statistical inference for metric space-valued data objects. We consider several important problems in statistical inference to show the utility of the quasi-DF.
 Note that the DF in Euclidean space has the correspondence theorem because its definition is closely relative to the Euclidean metric topology. Indeed, the DF is the Borel probability measure of the Cartesian products of left closed rays, which is a base of the Euclidean metric topology. While a metric space is equipped with a naturally metric topology that contains all open balls as a base,
 balls are generally not ordered, but concentric balls are. This ordering is essential for us to define the DFs in the same way as the ordered topology of one-dimensional Euclidean spaces, provided that we fix the center first. Using this center as the second variable, we can define the metric distribution function (MDF) in metric spaces as the counterpart of DF in Euclidean spaces.
(Figure~\ref{inference-framework}).

The rest of this article is organized as follows. We introduce the concepts of the MDF and the empirical MDF (EMDF) in Section~\ref{sec:MDF-EMDF},
and present their theoretical properties in Section~\ref{sec:theory}.
In Section~\ref{sec:statistical-method}, based on the MDF and the EMDF, we develop several nonparametric statistical inference procedures.
To demonstrate the MDF's effectiveness in practice, we employ the MDF-based methods on the synthetic and real-world datasets in
Sections~\ref{sec:monte-carlo-simulation} and \ref{sec:real-data}, respectively. Finally, we summarize our work for the MDF in Section~\ref{sec:summary}.
Technical proofs and some properties of EMDF are deferred to the ``Supplementary Material''.

\section{MDF and EMDF}\label{sec:MDF-EMDF}

\subsection{Notations}\label{metric}
An order pair $(\mathcal{M}, d)$ is called \emph{metric} space if $\mathcal{M}$ is a set and $d$ is a \emph{metric} or \emph{distance} on $\mathcal{M}$. Many spaces we have encountered are metric spaces. Examples include Euclidean space, Banach space, and connected Riemannian manifold. A metric space $(\mathcal{M}, d)$ is called \emph{separable} if it has a countable dense subset for the metric topology. A metric space $(\mathcal{M}, d)$ is said to be \emph{complete} if every Cauchy sequence converges in $\mathcal{M}$.  A completely separable metric
space is sometimes called a \emph{Polish} space.
Given a metric space $(\mathcal{M}, d)$, let $\bar{B}(u, r)=\{v: d(u,v) \leq r\}$ be the closed ball with the center $u$ and the radius $r \geq 0$, $B(u, r)=\{v: d(u,v) < r\}$ be the open ball and $\partial B(u,r)=\bar{B}(u, r)\setminus B(u, r)$ be the sphere.

If $(\mathcal{M}_k, d_k)$, $ k=1,  \ldots, K,$ are metric spaces, let $\mathcal{M}$ be the Cartesian product of $\mathcal{M}_k$,
denoted by $\prod\limits_{k=1}^K \mathcal{M}_k$. 
Here, Cartesian products of metric spaces are considered because they are useful for defining the independence measure in Section \ref{sec:statistical-method}.
For any $\mathbf{u}=(u_{1},\ldots ,u_{K})$ and $\mathbf{v}=(v_{1},\ldots ,v_{K})$ in $\mathcal{M}$, we can define a metric vector $e(\mathbf{u}, \mathbf{v})$ on the product space $\mathcal{M}$:
$$ e(\mathbf{u}, \mathbf{v})=\Big (d_{1}(u_{1},v_{1}),\ldots ,d_{K}(u_{K},v_{K})\Big ). $$
We also define $\bar{B}(\mathbf{u}, \mathbf{r}):=\prod_{k=1}^K \bar{B}(u_k, r_k) $ be the joint ball on the product space for a center vector $\mathbf{u} \in \mathcal{M}$ and a non-negative radius vector $\mathbf{r}=(r_1,\ldots, r_K)\succeq \mathbf{0}$.
For this product space, we can also assign a metric such that it is a metric space. For example, if we define
\begin{equation}\label{pmetric}
	d(\mathbf{u}, \mathbf{v})=\|e(\mathbf{u}, \mathbf{v})\|_p,
\end{equation} 
where $p\geq 1$ and $\|\cdot\|_p$ means the $\ell_p$ norm in $\mathbb{R}^K$. We can verify that $d(\cdot,\cdot)$ is a metric on $\mathcal{M}$.
Given a  point $\mathbf{u}\in\mathcal{M}$, $\pi_k(\cdot): \mathcal{M} \to \mathcal{M}_k$ is called the projection  on $\mathcal{M}_k$ if $\pi_k(\mathbf{u})=u_k$. For a set $A\subset\mathcal{M}$, we also define $\pi_k(A)=\bigcup\limits_{\mathbf{u}\in A}\{\pi_k(\mathbf{u})\}$.

Let $\mu$ be a (Borel) probability measure associated with an ordered $K$-tuple of random objects $\mathbf{X}=(X_1, \ldots, X_K)$ taking values in $\mathcal{M}(=\prod
\limits_{k=1}^K\mathcal{M}_k)$, and define $\mu \otimes \mu$ as the product measure on the measurable space.

\subsection{MDF and EMDF}\label{sec:mdf-emdf}
Denote the indicator function by $I(\cdot)$ and the radius vector $\mathbf{r}=e(\mathbf{u}, \mathbf{v})$ for $\mathbf{u}, \mathbf{v} \in \mathcal{M}$. We first define the metric distribution function (MDF) of $\mu$ on $\mathcal{M}$ that is the foundation of our proposed framework. For $\forall  \mathbf{u}, \mathbf{v} \in \mathcal{M}$, let
\begin{align*}
\delta(\mathbf{u},\mathbf{v},\mathbf{x})=& \prod\limits_{k=1}^K I\{x_{k}\in \bar{B}(u_k,r_k)\} =\prod\limits_{k=1}^K I\{x_{k}\in \bar{B}(u_k,d_k(u_k,v_k))\}.
\end{align*}
\begin{definition}
Given a probability measure $\mu$, we define the metric distribution function $ F^M_{\mu}(u, v)$  of  $\mu$ on $\mathcal{M}$: $\forall  \mathbf{u}, \mathbf{v} \in \mathcal{M}$,
	\begin{align*}
		F^M_{\mu}(\mathbf{u}, \mathbf{v})=\mu\left[\prod_{k=1}^K \bar{B}(u_k,r_k)\right]
		=E\left[\delta(\mathbf{u},\mathbf{v},\mathbf{X})\right].
	\end{align*}
\end{definition}

Suppose that $\{\mathbf{X}_1, \ldots,\mathbf{X}_n\}$ are $i.i.d.$ samples generated from a probability measure $\mu$ on a product metric space $\mathcal{M}=\prod\limits_{k=1}^K \mathcal{M}_k$.
We define the empirical metric distribution function (EMDF) associated with $\mu$ by the following formula naturally:
\begin{align*}
	F_{\mu,n}^M(\mathbf{u},\mathbf{v})=\frac{1}{n}\sum_{l=1}^n \delta(\mathbf{u},\mathbf{v},\mathbf{X}_l).
\end{align*}

\section{Theoretical analysis of MDF and EMDF}\label{sec:theory}
In this section, we first discuss some sufficient conditions for reconstructing probability measures from the MDFs and exhibit the properties of the convergence of the EMDFs. Additional properties of the EMDF are presented in the third part of the ``Supplementary Material."

\subsection{Fundamental reconstruction theorems of MDF}
Here we investigate whether a Borel probability measures $\mu$ on a separable metric space $(\mathcal{M},d)$ can be uniquely determined by the MDF $F^M_\mu(u,v)$.
We shall see that the answer depends on the probability measure and metric space. For separable metric spaces, \citet{federer2014geometric} introduced the following geometrical condition on the metric, named \emph{directionally $(\epsilon, \eta, L)$-limited}, to characterize the correspondence property of the MDF.

\begin{definition}[\citet{federer2014geometric}]\label{def:directionally-limited}
	A metric $d$ is called directionally $(\epsilon, \eta, L)$-limited at the subset $A$ of  $\mathcal{M}$, if $\epsilon>0$,
	$0<\eta\leq 1/3$, $L$ is a positive integer, and the following condition holds: if for each $a\in A$, $D\subseteq A\cap \bar{B}(a,\epsilon)$ such that $d(x,c)\geq \eta d(a,c)$ whenever $b,c\in D$ ($b\neq c$), $x\in \mathcal{M}$ with
	$$d(a,x)=d(a,c), d(x,b)=d(a,b)-d(a,x),$$
	then the cardinality of $D$ is no larger than $L$.
\end{definition}

\begin{figure}[htbp]
	\vspace*{-30pt}
	\begin{subfigure}{0.48\textwidth}
	    \centering
		\includegraphics[width=\textwidth]{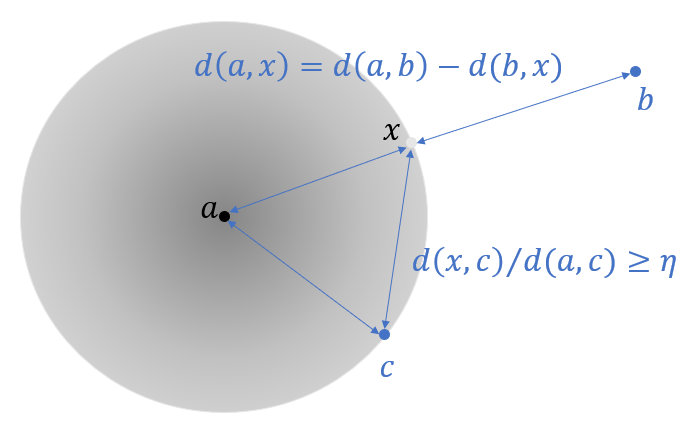}
    	\caption{} 
	\end{subfigure}
	\hfill
	\begin{subfigure}{0.48\textwidth}
	    \centering
	    \includegraphics[width=0.75\textwidth]{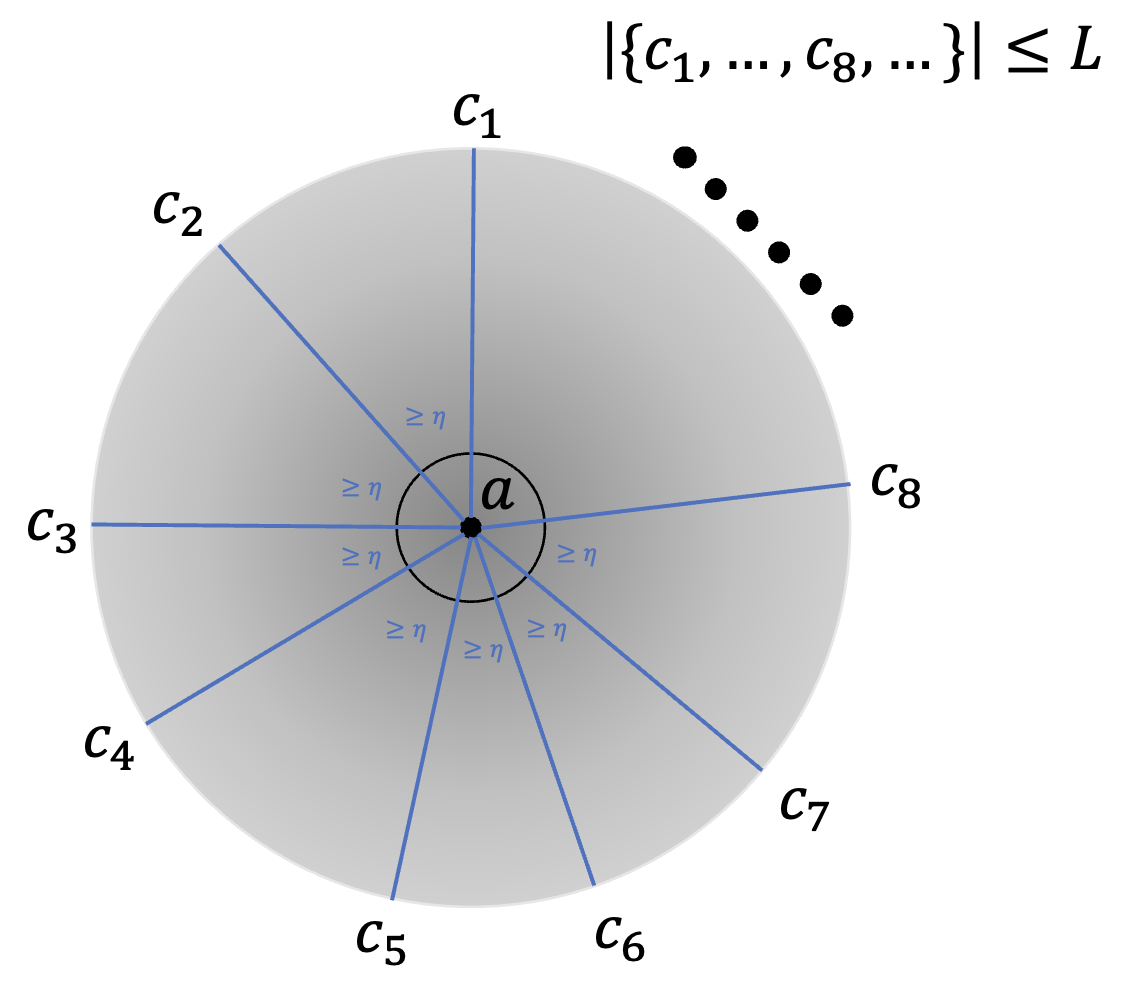}
	    \caption{}
	\end{subfigure}
	\caption{(a) Visualization of the direction in metric space by a 2-d Euclidean space example. (b) Visualization of the directionally $(\epsilon, \eta,  L)$-limited condition in the 2-d Euclidean space. For a given $\eta>0$, consider a circle $\mathcal{N}$ with the radius $r$ such that for any two points $c_i$ and $c_j$ in $\mathcal{N}$ we have  $d(c_i, c_j)/r\geq \eta$. The directionally $(\epsilon, \eta,  L)$-limited condition means that there exists an $L$ such that the cardinality of $\{c_1,\ldots,c_8,\ldots\}$ is always less than $L$.} \label{fig:directionally_limit}
	\vspace*{-15pt}
\end{figure}

Definition~2 ensures that the covering theorem holds; namely, given a ``thorough'' covering of a set by closed balls, there is a subcollection of pairwise disjoint balls that almost covers the set. The counterpart of this result in Euclidean spaces is the so-called Vitali covering theorem, which is important to the proof of the correspondence theorem in Euclidean spaces. Likewise, we need a similar condition for the correspondence theorem in metric spaces.  Figure \ref{fig:directionally_limit} intuitively illustrates this directionally limited assumption. Panel~(a) visualizes the definition of direction in metric space. The ratio of chord length and radius can measure the direction between two lines in a metric space. Panel~(b) illustrates the directionally limited assumption, which requires that the directions of every local point $a$ are finite. This concept of ``directionally $(\epsilon, \eta,  L)$-limited'' is essential to our reconstruction theory.  We examine a few examples to understand the implications of this condition.

First, if $(\mathcal{M}, \|\cdot\|)$  with the norm $\|\cdot\|$ is a Banach space, then the above definition implies
$$x=a+\frac{\|a-c\|}{\|a-b\|}(b-a),$$
thus $d(x,c)\geq \eta d(a,c)$ is equivalent to
$$\frac{d(x,c)}{d(a,c)}=\left\|\frac{b-a}{\|b-a\|}-\frac{c-a}{\|c-a\|}\right\|\geq \eta.$$
If $\mathcal{M}$ is a finite-dimensional Banach space, owing to the compactness of the unit sphere in $\mathcal{M}$, there exists a suitable $L$ for each $\eta>0$ such that the condition of directionally limited metric space holds.

Another case is when $\mathcal{M}$ is a finite-dimensional Riemannian manifold with bounded sectional curvature and $A$ is any compact subset of $\mathcal{M}$. Let $B(a,\epsilon)$ be a normal ball of $a\in A$,  and $\mbox{Exp}_a(\cdot)$ be the Riemannian exponential map and $\mbox{Log}_a(\cdot)$ be the Riemannian log map. By the bounded sectional curvature condition and Topogonov's theorem \citep{do1992riemannian}, 
we can find a universal constant $\lambda>0$ such that the for any 
 $b=\mbox{Exp}_a(\beta)$ and $c=\mbox{Exp}_a(\gamma)$ satisfying
\begin{align*}
	d(a,b)=&\|\beta\|_a, \ d(a,c)=\|\gamma\|_a, \ x=\mbox{Exp}_a \left[\frac{\|\gamma\|_a}{\|\beta\|_a}
 \beta \right], \textup{ and } d(x,c)/d(a,c)\geq \eta,
\end{align*}
 the inequality $\|\frac{\beta }{\|\beta\|_a} -\frac{\gamma }{\|\gamma\|_a}\|_a\geq \lambda\eta>0 $ holds. 
Thus, if we associate each $b\in D$ with the direction $\mbox{Log}_a(b) $, then by the compactness of unit sphere in the tangent space of $a$, there exists a suitable $L\in \mathbb{N}$ for each $\eta>0$.

The last but important case is when $(\mathcal{M},d)$ is  the metric space of a binary phylogenetic tree with
$\tau$ leaves,  where $\tau \in \mathbb{N}^{+}$ is fixed.
The space $\mathcal{M}$ is a Polish space and cubical complex \citep{lin2019total}.
Let $\mathcal{N}(\eta\epsilon; \bar{B}(x, \epsilon), d)$ be the $\eta\epsilon$-packing of $\bar{B}(x, \epsilon) \subseteq \mathcal{M}$
such that, for any $x, x^\prime \in \mathcal{N}(\eta\epsilon; \bar{B}(x, \epsilon), d)$,
the geodesic distance $d(x, x^\prime) \geq \eta\epsilon$.
Denote $\omega(\tau) = (2 \tau - 3) \times (2 \tau - 5) \times \cdots \times 3$,
for $\forall \eta \leq \frac{1}{3}$ and $\forall \epsilon \leq 1$, the space satisfies
\begin{align*}
	\sup_{x \in \mathcal{M}} card(\mathcal{N}(\eta \epsilon; \bar{B}(x, \epsilon), d)) \leq C \omega(\tau) \left( \frac{2}{\eta} \right)^{\psi(\tau)},
\end{align*}
for some constant $C < +\infty, \psi(\tau) < +\infty$.
This implies that the whole space is directionally-limited with $(\frac{1}{3}, 1, C\omega(\tau) 6^{\psi(\tau)})$.

Next, we provide an example of metric space that is not directionally limited. An infinite orthonormal base $B=\{e_1, e_2, \ldots\}$ in a separable Hilbert space $H$ is not directionally $(\epsilon, \eta,  L)$-limited. Let $a\in H$, and $b=a+e_i, c=a+e_j \in a+B$, then by the above discussion for Banach space, we have
$$x=a+\frac{\|e_j\|}{\|e_i\|}e_i \quad\textup{and}\quad \frac{d(x,c)}{d(a,c)}=\|e_j-e_i\|=\sqrt{2}\geq \eta.$$
for all $0<\eta\leq\frac{1}{3}$ and the cardinality of $a+B$ is infinite.

\begin{remark}\label{remark:directionally_limit}
Metric entropy and directionally $(\epsilon, \eta, L)$-limited are related concepts, and in certain classical metric spaces such as finite-dimensional Banach spaces and Riemannian manifolds with bounded sectional curvature, finite metric entropy implies directionally $(\epsilon, \eta, L)$-limited. Specifically, for a given $\delta$-covering number $N$, setting $\xi=\infty$, $\eta=2\delta$, and $\zeta=N$ guarantees that the directional limitability condition holds for any point $c$ and subset $B\subseteq A\setminus {a}$, as shown by the following inequality:
$$\frac{d(x,c)}{d(a,c)}=\left\|\frac{b-a}{\|b-a\|}-\frac{c-a}{\|c-a\|}\right\|\geq \eta=2\delta,$$
where $d$ denotes the metric, and $a,b,c,x$ are points in the metric space. Following the definition of metric entropy, the cardinality of $D$ is no larger than $N$. This result also holds for the Riemannian manifold with bounded sectional curvature. Let $\xi=\infty, \eta=2\delta\lambda,$ and $\zeta=N$, then the cardinality of $D$ is no larger than $N$ following inequality:
    \begin{align*}
    d(x,c)/d(a,c)\leq \lambda\vert \beta/\vert\beta\vert-\gamma/\vert\gamma\vert\vert.
    \end{align*}
    Thus, the finite-dimensional Banach space and Riemannian manifold with its usual metric are directionally $(\epsilon, \eta, L)$-limited if the metric entropy is finite for $\forall~\delta$.
\end{remark}

In a Euclidean space, two Borel probability measures $\mu=\nu$ if and only if their associated random objects $\mathbf{X}$ and $\mathbf{Y}$ share the common DF by the correspondence theorem \citep{halmos1956measure}. This correspondence lays the theoretical foundation for statistical inference. However,  DF depends on the linear structure and the order of real numbers. We do not have this structure in general metric space, and DF can no longer be defined. The following theorems delineate how MDF overcomes this major challenge.
Theorem \ref{SBall} shows that $d(u,\mathbf{X})$ and $d(u,\mathbf{Y})$ share the same DF for each location $u\in \mbox{supp}\{\mu\}$ if and only if $\mu=\nu$.

\begin{theorem}[The fundamental correspondence theorem of MDF]\label{SBall}
Denote $S=\{(u, v)\in\mathcal{M}\times\mathcal{M}:F^M_{\mu}(u,v)=F^M_{\nu}(u,v)\}$ for two given Borel probability measures, $\mu$ and $\nu,$ with their respective supports, $supp\{\mu\}$ and $supp\{\nu\},$ on $(\mathcal{M}, d)$. Suppose that $(\mathcal{M},d)$ is a Polish space and the metric $d$ is  directionally $(\epsilon, \eta, L)$-limited at $supp\{\mu\}$ and $supp\{\nu\}$, then $\mu\otimes\mu(S)=1$ (or $\nu\otimes\nu(S)=1$) if and only if $\mu=\nu$.
\end{theorem}
Theorem~\ref{SBall} ensures that the MDF has a one-to-one correspondence with a probability measure
when the metric is directionally $(\epsilon, \eta, L)$-limited at the support of the probability measure. The conditions of Theorem~\ref{SBall} may not be satisfied if $\mathcal{M}$ is a separable Hilbert space of infinite dimension. Corollary \ref{SBallcor} presents a reasonable condition so that the probability measure $\mu$ can still be determined by MDF in infinite dimension space.

\begin{corollary}\label{SBallcor}
For $\forall~\varepsilon>0$, suppose that there exists $\mathcal{M}_l\subseteq\mathcal{M}$ such that $\mu(\mathcal{M}_l)\geq1-\varepsilon$ (or $\nu(\mathcal{M}_l)\geq1-\varepsilon$) and the metric $d$ is directionally $(\epsilon(\mathcal{M}_l), \eta(\mathcal{M}_l), L(\mathcal{M}_l))$-limited at $\mathcal{M}_l$, then $\mu\otimes\mu(S)=1$ (or $\nu\otimes\nu(S)=1$) if and only if $\mu=\nu$.
\end{corollary}

Corollary \ref{SBallcor} includes separable Hilbert spaces as a special case.
For example, a random function, or a random curve $f(t)$ in a separable Banach space with unconditional Schauder base functions $\{\phi_j\}_{j=1}^\infty$, can be expanded as $f(t)=\sum_{j=1}^\infty \beta_j \phi_j(t),$ where the probability measure of the coefficients $\{\beta\}_{j=1}^\infty$, denoted as $\mu$, satisfies a sparse condition $\lim_{S\to\infty}\mu({\exists~j> S~,\beta_j\ne 0})=0$.
This setting is similar to the sparse priors used in \citet{Castillo2015Bayesian} and \citet{o2009review}, but it is important to note that we allow the underlying function space to be infinite-dimensional. In this example, the ``measure condition" of Corollary \ref{SBallcor} is satisfied. The metric condition of  Corollary \ref{SBallcor} implies that if $\mu$ and $\nu$ are two Borel probability measures on $\ell^2=\{(a_1,a_2,\ldots,a_i,\ldots):~\sum_i a_i^2<\infty\}$, and they share the common metric distribution function on any finite subspace of $\ell^2$, then we have $\mu=\nu$. According to the previous statement, many metric spaces, including the space of smooth functions, Riemannian manifold space, shape space, $\tau$-leaves binary phylogenetic spaces, 
satisfy the conditions of Corollary \ref{SBallcor}.

 In general, if the metric space is not linear, the geometric condition of ``directionally $(\epsilon, \eta,  L)$-limited'' cannot be induced from the compactness.
\citet{davies1971measures} gives a counter-example that there exists a compact metric space $(\mathcal{M},d)$ and two distinct Borel probability measures $\mu$ and $\nu$ on $\mathcal{M}$,
such that $\mu$ and $\nu$ agree on all closed balls.

Next, we extend the 1-1 correspondence theorem to product metric spaces. This extension is challenging because the topological structure of a product metric space may not be as simple as that of Euclidean space. For example, the product of two circles $S^1\times S^1$ is topologically not a sphere $S^2$ anymore.
Let $\mu$ and $\nu$ be two Borel probability measures on $\mathcal{M}=\prod\limits_{k=1}^K\mathcal{M}_k$, and
\begin{align*}
	\mathbf{M}_1=\{\mu:& \mu\textup{ is a discrete Borel probability measure on }\mathcal{M}\},\\
	\mathbf{M}_2=\{\mu:& \mu\textup{ is a  Borel probability measure on }\mathcal{M}  \textup{ such that }\\
	&\forall \mathbf{u}\in\mathcal{M} \textup{ and } \mathbf{v} \sim\mu,  e(\mathbf{u},\mathbf{v}) \textup{ has continuous density function}\}.
\end{align*}
We have the following fundamental correspondence theorem of the joint metric distribution function in a product metric space.
\begin{theorem}[The fundamental correspondence theorem of joint MDF]\label{JBall}
Given two Borel probability measures, $\mu$ and $\nu,$ on a product Polish space $\mathcal{M}=\prod_{k=1}^K \mathcal{M}_k $, let $\mathbf{S}=\{(\mathbf{u},\mathbf{v})\in\mathcal{M}\times\mathcal{M}:F^M_{\mu}(\mathbf{u}, \mathbf{v})=F^M_{\nu}(\mathbf{u}, \mathbf{v})\}$. Suppose that $d_k$ is directionally-$(\xi_k,\eta_k,L_k)$ limited at $\mathcal{M}_k$ and $\mu = \alpha\eta+(1-\alpha)\gamma$ for some $\alpha\in[0,1]$, $\eta\in \mathbf{M}_1 $ and $\gamma\in\mathbf{M}_2$, then  $\mu\otimes\mu(\mathbf{S})=1$ (or $\nu\otimes\nu(\mathbf{S})=1$) if and only if  $\mu=\nu$.
\end{theorem}
Similar to Corollary \ref{SBallcor}, we have the following corollary for the product space.

\begin{corollary}\label{JBallcor}
For $\forall~\varepsilon>0$, suppose that there exists $\mathcal{M}_l\subseteq\mathcal{M}$ such that $\mu(\mathcal{M}_l)\geq1-\varepsilon$ (or $\nu(\mathcal{M}_l)\geq1-\varepsilon$) and the metric $d_k$ is directionally $\big(\epsilon(\pi_k(\mathcal{M}_l)), \eta(\pi_k(\mathcal{M}_l)), L(\pi_k(\mathcal{M}_l))\big)$-limited at $\pi_k(\mathcal{M}_l)$. If the combination $\mu=\alpha \eta+(1-\alpha) \gamma$ for $\eta \in M_{1}$ and $\eta \in M_{2}$, then $\mu\otimes\mu(\mathbf{S})=1$ (or $\nu\otimes\nu(\mathbf{S})=1$) if and only if $\mu=\nu$.
\end{corollary}

\subsection{Main properties of EMDF}
Here, we provide EMDF's Glivenko-Cantelli property and Donsker property.
First, we define the collection of the indicator functions of closed balls on $\mathcal{M}$:
\begin{align*}
	\mathcal{F}=\{ \delta(\mathbf{u},\mathbf{v}, \cdot): \mathbf{u}\in\mathcal{M} \textup{ and }\mathbf{v} \in\mathcal{M} \}.
\end{align*}
The uniform convergence property of EMDF is given as follows.
\begin{theorem}[The Glivenko-Cantelli type property of EMDF]\label{uniform1}
	Let $\mathcal{M}=\prod\limits_{k=1}^K \mathcal{M}_k$ be a product space and $\mu$ be a probability measure on it.
	Suppose that $\{\mathbf{X}_1, \ldots,\mathbf{X}_n\}$ is a sample of i.i.d observations from $\mu$. Define $\mathcal{F}(\mathbf{X}_1^n):=\{(f(\mathbf{X}_1),\ldots,f(\mathbf{X}_n)) | f\in\mathcal{F} \}$.   If  $\mu$ satisfies that
	$$\frac{1}{n}E_\mathbf{X}\left[\log (card(\mathcal{F}(\mathbf{X}_1^n)))   \right]\to 0,$$
	where $card(\cdot)$ is the cardinality of  a set,  we have
	the Glivenko-Cantelli property of our empirical metric distribution function:
	$$\lim_{n\rightarrow\infty}\sup_{\mathbf{u}\in \mathcal{M}, \mathbf{v} \in \mathcal{M}} |  F_{\mu,n}^M(\mathbf{u},\mathbf{v})-F^M_{\mu}(\mathbf{u}, \mathbf{v})|=0,\ a.s.$$
\end{theorem}
The conditions of Theorem~\ref{uniform1} are often satisfied in practice.
The first example is $\mathcal{M}=\mathbb{R}^q$ with the $\ell_p$-norm (where $p$ is a positive integer or $\infty$), and $\mu$ is an arbitrary probability measure because the set of $\ell_p$ ball has a finite VC-dimension. We also allow the dimension of $\mathcal{M}$ to increase as the sample size increases if $\mathcal{M}$ is a Euclidean space.
Since the VC-dimension of closed balls in Euclidean space $R^{q}$ is $q+2$ (see Example 4.14 in \cite{wainwright_2019}), if $q=o(\frac{n}{\log n })$ the Glivenko-Cantelli property still holds (Lemma 4.14  in \cite{wainwright_2019}).
The second example is that $\mathcal{M}$ is a smooth regular curve in Euclidean space or a sphere in $\mathbb{R}^q$ with the geodesic distance, and $\mu$ is an arbitrary probability measure. In this case, we can reparametrize $\mathcal{M}$ to be a unit speed curve such that every geodesic ball in $\mathcal{M}$ is mapped to an interval in $\mathbb{R}$ and the set of intervals in $\mathbb{R}$ has a finite VC-dimension.
The third example is that  $\mathcal{M}$ is a set of polygonal curves in $\mathbb{R}^d$ with the Hausdorff distance for the Fr{\'e}chet distance  \citep{driemel2019vc}  and $\mu$ is an arbitrary probability measure.
Another example is that $\mathcal{M}$ is a separable Hilbert space with a probability measure $\mu$ with support on a finite-dimensional subspace because  the set of balls on the support of $\mu$ has a finite VC-dimension.

Based on the two reconstruction theorems, whether the two probability measures are identical depends on whether their MDFs are the same over their support sets but not the whole space. This leads us to consider the Glivenko-Cantelli type property for the MDF over the sample set because the sample set contains the information that supports the underlying unknown probability measure.
\begin{corollary}[A concentration inequality of EMDF]\label{uniform2}
	Let $\mathcal{M}=\prod\limits_{k=1}^K \mathcal{M}_k$ be a product space and $\mu$ be a probability measure on it. For each $t>0$, there exists a universal constant $N(t)\in\mathbb{N}$
	such that for all $n\geq N(t)$, we have
	\begin{align*}
		P(\max\limits_{1\leq i,j\leq n} |F_{\mu,n}^M(\mathbf{X}_i,\mathbf{X}_j)-F_\mu^M(\mathbf{X}_i,\mathbf{X}_j)|>t)\leq 2n\exp({-\frac{nt^2}{32} }),
	\end{align*}
	which leads to
	\begin{align*}
		\max\limits_{1\leq i,j\leq n}|F_{\mu,n}^M(\mathbf{X}_i,\mathbf{X}_j)-F_\mu^M(\mathbf{X}_i,\mathbf{X}_j)| \stackrel{a.s.}{\longrightarrow} 0,~\textup{as}~n \rightarrow \infty.
	\end{align*}
\end{corollary}
Without restriction on metric spaces and probability measures, Theorem \ref{uniform2} shows that the EMDF has the concentration phenomenon at an exponential convergence rate for a sufficiently large sample. This uniform convergence result over the sample set is essential when we apply the EMDF to analyze data objects in metric spaces as it is the data analysis in a Euclidean space.
The other important convergence property of the EMDF is the convergence in distribution, called the Donsker property, which is similar to the central limits theorem.
\begin{theorem}[The Convergence of  Metric Distribution Process]\label{bep}
	Let $\mathcal{M}=\prod\limits_{k=1}^K \mathcal{M}_k$ be a product space and $\mu$ be a probability measure on it. Define
\begin{align*}
	\mathbb{G}_n(\mathbf{u},\mathbf{v}) =\sqrt{n}(  F_{\mu,n}^M(\mathbf{u},\mathbf{v})-F^M_{\mu}(\mathbf{u}, \mathbf{v})),~  \mathbf{u},\mathbf{v}\in\mathcal{M}.
\end{align*}
	If $\mathcal{F}$ is a VC class with VC-dimension  $V(\mathcal{F})<\infty$, then
	we have the Donsker property of the metric distribution process: $
	\{{\mathbb{G}_n(\mathbf{u},\mathbf{v}): \mathbf{u},\mathbf{v}\in\mathcal{M} }\}$
	converges in distribution to a Gaussian process $\{\mathbb{G}_\mu(\mathbf{u},\mathbf{v}), \mathbf{u}\in\mathcal{M} \textup{ and }\mathbf{v} \in\mathcal{M}\}$, with zero  mean and the covariance function:
	\begin{align*}
		E \mathbb{G}_\mu(\mathbf{u}_1,\mathbf{v}_1  ) \mathbb{G}_\mu(\mathbf{u}_2,\mathbf{v}_2 )=   &\mu\left( \bar{B}(\mathbf{u}_1, e( \mathbf{u}_1,  \mathbf{v}_1)) \cap \bar{B}(\mathbf{u}_1, e( \mathbf{u}_2,  \mathbf{v}_2))\right )
		- F^M_{\mu}(\mathbf{u}_1, \mathbf{v}_1)F^M_{\mu}(\mathbf{u}_2, \mathbf{v}_2).
	\end{align*}
\end{theorem}
The conditions in Theorem~\ref{bep} also imply the Glivenko-Cantelli property of the EMDF.
It is noteworthy that EMDF has  the Glivenko-Cantelli and Donkser properties in the infinite-dimensional cases if we impose some entropy and continuity conditions on the probability measures:
\begin{corollary}
     Assume  the following conditions hold: 
    \begin{enumerate}
        \item  $\int_0^\infty \sqrt{\frac{\log N(\epsilon,\mathcal{M},t)}{t}}dt<\infty $, where $N(\epsilon,\mathcal{M},t)$ is the covering number.
        \item   The CDF of $d(\mathbf{u},\mathbf{X})$,  has probability density  function $f(\mathbf{u},\mathbf{r})$ for all $\mathbf{u}\in\mathcal{M} $, and satisfies $\sup_{ \mathbf{u},\mathbf{r}}f(\mathbf{u},\mathbf{r})\leq C_f<\infty $ for some constant $C_f$. 
    \end{enumerate}
Then, both Glivenko-Cantelli and Donsker properties hold.
\end{corollary}

\begin{remark}
The preceding result can be extended without significant additional effort to replace $\mathcal{M}_k$ in the first condition with the projection of the support of $\mu$ onto $\mathcal{M}_k$. The first condition  on $\mathcal{M}_k$ is mild and can be satisfied by any space satisfying $\log N(t,\mathcal{M},d)=O(t^{-c})$ for some constant $c<1$, including 
\begin{enumerate}
\item Bounded subsets  of the $\alpha$-times continuously differentiable function space $ C_1^\alpha(\mathcal{X})$ defined on $\mathcal{X}$ equipped with the $\|\cdot\|_\infty$-norm or $L_r(Q)$- norm for a certain probability measure $Q$ on $\mathcal{X}$. Here, $\mathcal{X}$ is a bounded convex subset of $\mathbb{R}^p$ and $\alpha>d$, as proved by  Theorem 2.7.1 in \cite{wellner2013weak}.
    \item Bounded subsets of Riemannian manifold with bounded sectional curvature by the  Bishop-Gromov packing lemma \citep{petersen2006riemannian}, such as a bounded subset of SPD matrices manifold equipped with the affine-invariant metric and $p$-dimensional sphere.
     \item Bounded subsets or balls of $\mathbb{R}^p$ equipped with $\|\cdot\|_q$-norm for $1\leq q\leq \infty$ \citep{wainwright_2019}.
     \item Bounded subsets  of the binary phylogenetic tree with
$\tau$ leaves,  where $\tau \in \mathbb{N}^{+}$ is fixed.

\end{enumerate}
\end{remark}

\section{MDF based statistical methods}\label{sec:statistical-method}
In this section, we discuss using the MDF to conduct statistical inference in a few important and common problems.

\subsection{Homogeneity test}\label{sec:homogeneity-test}
A common and basic hypothesis testing problem in statistical inference is whether samples are generated from the same distribution.
Suppose we have data objects from unknown Borel probability measures, $\mu_1, \mu_2, \ldots, \mu_K,$ on a metric space $(\mathcal{M},d)$
and need to check whether they are homogeneous,
i.e., testing $H_{0}: \mu_1 = \mu_2 = \cdots = \mu_K$.

We introduce a homogeneity measure based on MDF, called metric Cram{\'e}r-von Mises (MCVM). Let $F^M_{\mu_k}(\mathbf{u}, \mathbf{v})$ be the MDFs for $\mu_k$, $\mu$ is the mixture distribution of $\mu_1, \ldots, \mu_K$ with proportions $p_1, \ldots, p_K$, and $F^M_{\mu}(\mathbf{u}, \mathbf{v})$ be the MDF of $\mu$, we use some Cram{\'e}r-von Mises-type criteria to evaluate the distinction of $F^M_{\mu_k}(\mathbf{u}, \mathbf{v})$ and $F^M_\mu(\mathbf{u}, \mathbf{v})$ for at $\mathbf{u}, \mathbf{v}$ from $\mu_1, \ldots, \mu_K$:
\begin{align*}
	\textup{MCVM}(\mu_k \| \mu) = \int_{(\mathbf{u}, \mathbf{v}) \in \mathcal{M} \times \mathcal{M}}  p^2_k w(\mathbf{u}, \mathbf{v}) \left(F^M_{\mu_k}(\mathbf{u}, \mathbf{v}) - F^M_{\mu}(\mathbf{u}, \mathbf{v})\right)^2 d\mu_k(\mathbf{u})d\mu_k(\mathbf{v}),
\end{align*}
where $w(\mathbf{u}, \mathbf{v}) = \exp\{ -\frac{(d(\mathbf{u}, \mathbf{v}))^2}{2\sigma^2} \}$.
We aggregate $\textup{MCVM}(\mu_k \| \mu)$ by defining
\begin{align*}
	\textup{MCVM}(\mu_1, \ldots, \mu_K) = \sum_{k=1}^K p_k^2\textup{MCVM}(\mu_k \| \mu).
\end{align*}

For each $k = 1, \ldots, K$, let $\mathcal{X}_k$ be the $k$-th sample set of $\mathbf{X}^{(k)}_1, \ldots, \mathbf{X}^{(k)}_{n_k} \overset{i.i.d}{\sim} \mu_k$. Then
$\textup{MCVM}(\mu_k \| \mu)$ can be estimated on the basis of EMDF:
\begin{align*}
	\widehat{\textup{MCVM}}(\mu_k \| \mu) &=
	\frac{1}{n_k^2} \sum_{\mathbf{X}^{(k)}_i, \mathbf{X}^{(k)}_j \in \mathcal{X}_k} (\hat{p}_k)^2 w(\mathbf{X}^{(k)}_i, \mathbf{X}^{(k)}_{j}) \left( F^M_{\mu_k, n_k}(\mathbf{X}^{(k)}_i, \mathbf{X}^{(k)}_j) - F_{\mu, n}^M(\mathbf{X}^{(k)}_i, \mathbf{X}^{(k)}_j)\right)^2, 
\end{align*}
where $\hat{p}_k = n_k / \sum_{k=1}^K n_k$. Thus, $\widehat{\textup{MCVM}}(\mu_1, \ldots, \mu_K) = \sum_{k=1}^K (\hat{p}_k)^2 \widehat{\textup{MCVM}}(\mu_k \| \mu)$. We use the median heuristic for choosing the $\sigma^2$ as the median of $\{ d(\mathbf{X}, \mathbf{X}'): \mathbf{X}, \mathbf{X}' \in \cup_{k=1}^K \mathcal{X}_k \}$.

Following the above Theorems, we can obtain the theoretical properties of MCVM. 
\begin{proposition}\label{prop:homogeneity}
\begin{enumerate}
    \item[(a)] Suppose that the conditions of Theorem \ref{SBall} or Corollary \ref{SBallcor} hold and $p_k>0, k=1,\ldots, K$, then $\textup{MCVM}(\mu_1, \ldots, \mu_K)=0$ if and only if $\mu_1 = \mu_2 = \cdots = \mu_K$.
    \item[(b)] Suppose that the conditions of Theorem \ref{uniform1} hold and $p_k>0, k=1,\ldots, K$, then $$\widehat{\textup{MCVM}}(\mu_1, \ldots, \mu_K)\xrightarrow[n_1,\ldots,n_K\rightarrow\infty]{a.s.}\textup{MCVM}(\mu_1, \ldots, \mu_K).$$
    \item[(c)] Suppose $p_k>0\; (k=1,\ldots, K)$, under the conditions of Theorem \ref{uniform2}, when the null hypothesis holds,  $$n\widehat{\textup{MCVM}}(\mu_1, \ldots, \mu_K)\xrightarrow[n_1,\ldots,n_K\rightarrow\infty]{d}\sum_{l=1}^\infty\lambda_lZ_l^2,$$
    where $Z_l$ are i.i.d. standard normal random variables and $\lambda_l$ are the constants depending on $\mu_1, \ldots, \mu_K$, $l=1, \ldots, \infty.$ $n\widehat{\textup{MCVM}}(\mu_1, \ldots, \mu_K)$ can serve as a test statistic for homogeneity, which is consistent against the alternatives.
    \end{enumerate}
\end{proposition}
To test $H_0$, we can use permutation to approximate the $p$-value directly. On the other hand, the asymptotic distribution of MCVM in Proposition~\ref{prop:homogeneity} (c) motivates us to test homogeneity by estimating $\{\lambda_i\}_{i\in \mathbb{N}}$ when $n$ is sufficiently large. We name this test procedure as a spectrum-based test and study its numerical performance in Section~\ref{sec:comparison-large-sample-regime}. We defer its implementation details and theoretical property in Section~4 of Supplementary Materials.  

\subsection{Mutual independence test}\label{sec:independence-test}
Another fundamental problem in statistical inference is testing the mutual independence among several elements of a random object. 
Suppose $\mathbf{X}=(X_1, \ldots, X_K)$ is a random object of $K$-tuple random objects ($K \geq 2$) on a metric space $(\mathcal{M}, d)$,
in which $\mathbf{X}$ is associated with probability measure $\mu$,
and $X_k$ is associated with probability measure $\mu_k$ on $\mathcal{M}_k$ for $k=1, \ldots, K$.
The study of mutual independence is formulated as testing $H_0: \mu = \mu_1 \otimes \cdots \otimes \mu_K$.

It is very convenient to utilize the MDF to measure mutual dependence because of the definition of the MDF in product metric spaces.
Following Hoeffding's dependence paradigm \citep{hoeffding1948independence}, we
integrate the difference between the joint MDF associated with $\mathbf{X}$ and
the product of marginal MDFs associated with $X_k$'s. We then obtain our metric association (MA) measure:
\begin{align*}
	\mathrm{MA}(\mu, \otimes_{k=1}^K\mu_k)
	= &\int \left(F^M_\mu(\mathbf{u},\mathbf{v})-\prod_{k=1}^KF^M_{\mu_k}(u_k, v_k)) \right)^2d\mu(\mathbf{u})d\mu(\mathbf{v}).
\end{align*}
When $K=2$, $\mathrm{MA}(\mu, \otimes_{k=1}^2\mu_k)$ is the square of ball covariance in \citet{pan2020ball}. When the entries of $\mathbf{X}$ are dependent, then Theorem \ref{JBall} implies that $\mathrm{MA}(\mu, \otimes_{k=1}^K\mu_k)>0$.
Suppose that $\mathbf{X}_1, \ldots, \mathbf{X}_n$ are $i.i.d.$ observations of $\mathbf{X}$ associated with
the Borel probability measures $\mu$.
The consistent estimator for $\mathrm{MA}(\mu, \otimes_{k=1}^K\mu_k)$ is given by
\begin{align*}
	\widehat{\mathrm{MA}}(\mu, \otimes_{k=1}^K\mu_k)
	=\frac{1}{n^{2}}\sum_{i,j=1}^{n}\Big( F_{\mu,n}^M(\mathbf{X}_i,\mathbf{X}_j) -\prod_{k=1}^K F^M_{\mu_k,n} (X_{ik}, X_{jk})\Big)^{2}.
\end{align*}

The following proposition demonstrates the practical use of Theorems~\ref{SBall}-\ref{uniform2}.
\begin{proposition}\label{prop:independence}
    \begin{enumerate}[(a)]
        \item Under the conditions in Theorem~\ref{JBall} or Corollary~\ref{JBallcor}, $\mathrm{MA}(\mu, \otimes_{k=1}^K\mu_k)=0$ if and only if $\mu=\otimes_{k=1}^K\mu_k$.
        \item Under the conditions in Theorem~\ref{uniform1}, we have $\widehat{\mathrm{MA}}(\mu, \otimes_{k=1}^K\mu_k) \xrightarrow[n\rightarrow\infty]{a.s.} \mathrm{MA}(\mu, \otimes_{k=1}^K\mu_k).$
        \item Suppose the conditions in Theorem~\ref{uniform1} hold, then under the null hypothesis
        $$n\widehat{\mathrm{MA}}(\mu, \otimes_{k=1}^K\mu_k)\xrightarrow[n\rightarrow\infty]{d}\sum_{l=1}^{\infty}\lambda_{l}Z_{l}^2,$$ 
        where $\{Z_l\}_{l\in\mathbb{N}}$ is a countable sequence of $i.i.d.$ standard normal random variables and $\{\lambda_l\}_{l\in \mathbb{N}}$ is a descending ordered. Thus, $\widehat{\mathrm{MA}}(\mu, \otimes_{k=1}^K\mu_k)$ is a consistent test for any fixed alternative hypothesis. 
    \end{enumerate}
\end{proposition}
Motivated by Proposition~\ref{prop:independence}(c), we can derive an estimator for the $\{\lambda_{l}\}_{l\in \mathbb{N}}$ and give a spectrum-based test by following the similar procedure for MCVM (See Section 4 in Supplementary Materials).
Also, we can approximate $p$-values by permutation when the sample size is relatively small. 
The numerical comparison between the permutation-based and spectrum-based mutual independence test is conducted in Section~\ref{sec:comparison-large-sample-regime}.

\section{Monte Carlo Studies}\label{sec:monte-carlo-simulation}

\subsection{Consistency of tests: large-sample regime}\label{sec:comparison-large-sample-regime}
We investigate the consistency of the permutation-based and spectrum-based tests proposed above. 
We simulate datasets drawn from the multivariate Gaussian distribution $N(\mu, I_{2\times 2})$,
the von-Miser Fisher distribution $\mathcal{V}(w)$ with concentration parameter $\|w \|$,
and the Wishart distribution $\mathcal{W}(\Sigma)$ with degree of freedom 8. They are common distributions in Euclidean, spherical, and SPD matrices space.
Here, we vary sample size $n$ from 200 to 600.
The significance level is fixed at 0.05.

For the assessment of the homogeneity test, we design the following models:
\begin{itemize}
  \item {Euclidean:} $X \sim N(\textbf{0}, I_{2\times 2})$, $Y \sim N(\mu^y, I_{2\times 2})$, $Z \sim N(\mu^z, I_{2\times 2})$.
  \item {Sphere:} Let $X \sim \mathcal{V}((1, 0)^\top)$, $Y \sim \mathcal{V}(w^y)$ and $Z \sim \mathcal{V}(w^z)$. 
  \item {SPD:} $X \sim \mathcal{W}(I_{2\times 2}), Y \sim \mathcal{W}(\Sigma^y), Z \sim \mathcal{W}(\Sigma^z)$, where $\Sigma^y, \Sigma^z$ are $2 \times 2$ matrices.
\end{itemize}
We set $\mu^y = \mu^z = \textbf{0}$, $w^y = w^z = (1, 0)^\top$, and $\Sigma^y = \Sigma^z = I_{2\times 2}$ to examine Type-I errors for the above settings, respectively. 
To check the consistency of the proposed tests, we set $\mu^y = (0.2, 0.2)^\top, \mu^z = (-0.2, -0.2)^\top$, 
$w^y = \frac{1}{\sqrt{5}}(2, 1)^\top, w^z = \frac{1}{\sqrt{2}}(1, 1)^\top$, 
$\Sigma^y_{ij} = 0.1^{|i-j|}$, $\Sigma^z_{ij} = (-0.1)^{|i-j|}$.

To detect the mutual dependence among $(X, Y, Z)$, we consider the three models below.
\begin{itemize}
  \item {Euclidean:} $(X, Y, Z) \sim N(\mathbf{0}_3, \Sigma)$.
  \item {Spherical:} $X^\prime, Y^\prime, Z^\prime$ are random variables in $\mathbb{R}^2$ and $(X^\prime, Y^\prime, Z^\prime) \sim N(\mathbf{0}_6, \Sigma)$. And $X \sim \mathcal{V}(X^\prime)$, $Y \sim \mathcal{V}(Y^\prime)$, $Z \sim \mathcal{V}(Z^\prime)$.
  \item {SPD:} Draw $(X^\prime, Y^\prime, Z^\prime)$ from $N(\mathbf{0}, \Sigma)$. Let $\mathcal{W}(a)$ be a $2\times 2$ matrix with diagonal value 10 and non-diagonal value $a$, we set $X \sim \mathcal{W}(X^\prime), Y \sim \mathcal{W}(Y^\prime), Z \sim \mathcal{W}(Z^\prime)$.
\end{itemize}
We set $\Sigma$ as the identity matrix to assess Type-I errors. As for power analysis, we set $\Sigma_{ij} = 0.2^{|i-j|}$ for Euclidean data and $\Sigma_{ij} = 0.6^{|i-j|}$ for spherical and SPD datasets. 
We now introduce the distance measure. We employ Euclidean/geodesic distance for data in Euclidean/spherical space.
We utilize the Cholesky distance \citep{dryden2009noneuclidean} to measure the difference between two SPD matrices $P_1, P_2$,
which is defined as $\| \textup{chol}(P_1) - \textup{chol}(P_2) \|_F$, where $\| \cdot \|_F$ is Frobenius norm
and $\textup{chol}(\cdot)$ is the Cholesky decomposition.

\begin{figure}[t]
        \vspace{-30pt}
	\includegraphics[width=\linewidth]{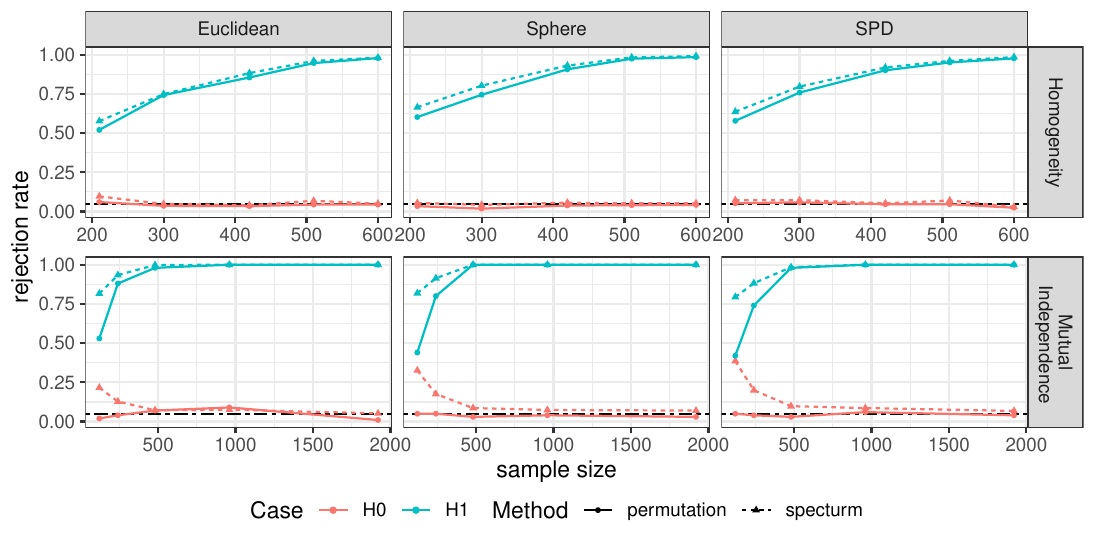}
	\vspace{-25pt}
	\caption{Rejection rate of the proposed homogeneity and mutual independence tests. The line type distinguishes permutation-based and spectrum-based tests; the color distinguishes the $H_0$ and $H_1$. The black dashed line is the significance level.}
	\label{fig:large_homogeneity}
\end{figure}

Figure~\ref{fig:large_homogeneity} displays the rejection rate of the homogeneity and mutual-independence tests under 500 Monte Carlo runs. When the null hypotheses hold, both permutation and spectrum tests control the rejection rate well around the significance level. When $n$ is not sufficiently large, the spectrum-based test may have an excessive rejection rate than the permutation test. Besides, controlling the type-I error of the mutual independence test requires more samples than the homogeneity test. This is because the form of $h_2$ involves multiple terms depending on the unknown probability measure $\{\mu_{k}\}_{k=1}^K$, implying more samples are required to control the approximation error of $h_2$. When the alternative hypotheses hold, the empirical powers of the permutation and spectrum tests are close, and they both increase to 1 as $n$ goes to infinity, reflecting the consistency of the two tests.

\subsection{Power analysis: finite-sample regime}

We first depict the setting for testing the homogeneity of $X, Y, Z$. Let $\Sigma{(a)}$ be a $p$-by-$p$ matrix whose non-diagonal entries are $a$, and diagonal entries are 1.
\begin{itemize}
    \item Euclidean: (i) $X \sim N(\mathbf{0}, I_{2\times 2})$, $Y \sim N(\kappa \times \mathbf{1}, I_{2\times 2})$, $Z \sim N(-\kappa \times \mathbf{1}, I_{2\times 2})$; (ii) 
    $X, Y, Z$ are drawn from zero-mean multivariate $t$-distributions with degree of freedom 3, $3-\kappa$, and $3+\kappa$.
    \item Spherical: (i) $X, Y, Z$ comes from the von Miser-Fisher distributions with concentration parameter 2.5, with directions $(\cos{\frac{\pi}{4}}, \sin{\frac{\pi}{4}})^\top$, $(\cos{(\frac{\pi}{4}(1 + \tanh{\kappa}))}, \sin{(\frac{\pi}{4}(1 + \tanh{\kappa}))})^\top$, and $(\cos{(\frac{\pi}{4}(1 - \tanh{\kappa}))}, \sin{(\frac{\pi}{4}(1 - \tanh{\kappa}))})^\top$, respectively. (ii) $X$ is the equiv-probability mixture of von-Miser Fisher distributions with directions $(\cos{\frac{\pi}{4}}, -\sin{\frac{\pi}{4}})^\top, (\cos{\frac{\pi}{4}}, \sin{\frac{\pi}{4}})^\top$ and concentration parameter 2.5. $Y, Z$ are the same as $X$ except that directions are replaced with $(\cos{\alpha}, \sin{\alpha})^\top, (\cos{\alpha}, -\sin{\alpha})^\top$ and $(\cos{\beta}, \sin{\beta})^\top, (\cos{\beta}, -\sin{\beta})^\top$, respectively.
    Here, $\alpha = \frac{\pi}{4}(1 + \tanh{\kappa})$ and $\beta = \frac{\pi}{4}(1 - \tanh{\kappa})$.
    \item SPD matrix: (i) $X$, $Y$, $Z$ are drawn from the Wishart distributions with the degree of freedom 8. The scale matrices of $X, Y, Z$ are $I_{3\times 3}$, $\Sigma_{3}(\kappa)$, $\Sigma_{3}(-\kappa)$; (ii) $X$, $Y$, $Z$ are the Wishart distributions with the degree of freedom $8-\kappa$, 8, $8+\kappa$, respectively. The scale matrices of $X$, $Y$, $Z$ are $\frac{8}{8-\kappa} \times \Sigma_3(0.1)$, $\Sigma_3(0.1)$, and $\frac{8}{8+\kappa} \times \Sigma_3(0.1)$.
\end{itemize}
For each metric space, case (i) makes distributions only have a difference on Fr{\'e}chet mean, while case (ii) only has a difference on Fr{\'e}chet variance. 

Next, we describe the setting for testing the mutual independence among $(X, Y, Z)$.   
\begin{itemize}
    \item Euclidean: (i) $(X, Y, Z) \sim N(\mathbf{0}, \Sigma_3(\kappa))$; (ii) $(X, Y, Z)$ follows the equiv-probability mixture of $N(\mathbf{0}, \Sigma_3(\kappa))$ and $N(\mathbf{0}, \Sigma_3(-\kappa))$.
    \item Spherical: first sample $(X', Y', Z')$ from $N(\mathbf{0}, \Sigma_3(\kappa))$. (i) $X, Y, Z$ come from three von-Miser distributions with concentration parameters 2.8 and directions: $(\cos(X'), \sin(X'), 0, 0)^\top$, $(\cos(Y'), \sin(Y'), 0, 0)$ and $(\cos(Z'), \sin(Z'), 0, 0)$, respectively. (ii) $X, Y, Z$ come from three von-Miser distributions whose directions are $(1, 0, 0, 0)^\top$ and concentration parameters $|X'|, |Y'|, |Z'|$, respectively. 
    \item SPD matrix: $(X', Y', Z') \sim N(\mathbf{0}, \Sigma_3(\kappa))$, then we generate $X, Y, Z$ from three Wishart distributions with parameters: (i) scale matrix $\Sigma_3(0.1)$ and degree-of-freedom are $3+8|X'|$, $3+8|Y'|$ and $3+8|Z'|$; or (ii) scale matrices $\frac{1}{3+12|X'|} \times \Sigma_3(0.1)$, $\frac{1}{3+12|Y'|} \times \Sigma_3(0.1)$, $\frac{1}{3+12|Z'|} \times \Sigma_3(0.1)$ and degree-of-freedom are $3+12|X'|$, $3+12|Y'|$, $3+12|Z'|$.
\end{itemize}
For each metric space, $X, Y, Z$ have mean dependence in (i) and variance dependence in (ii). 

\begin{figure}[htbp]
    \includegraphics[width=\linewidth]{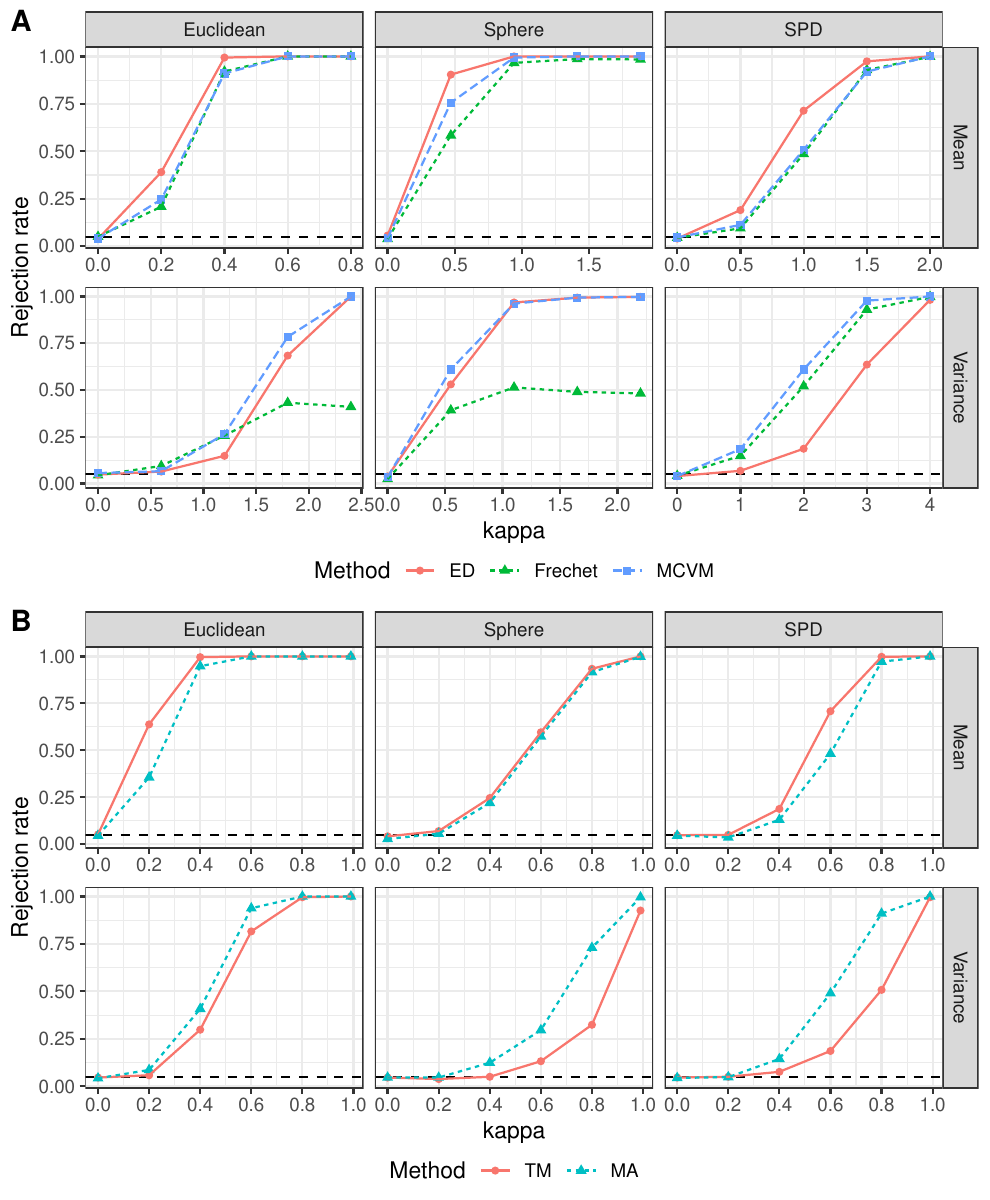}
	\vspace{-25pt}
	\caption{Rejection rate of hypothesis tests when $\kappa$ increases. A: the homogeneity test. B: the mutual independence test. Methods are distinguished by point and line. The black dashed line is the nominal significance level.}
	\label{fig:finite_kappa_change}
\end{figure}

We study the empirical power of the proposed tests when the distribution discrepancy/dependence strength $\kappa$ varies, but $n$ is fixed.
We compare our proposed method with energy distance (ED, \citet{szekely2004testing}), 
Fr{\'e}chet variance analysis (FVA, \citet{muller2019frechet}) for homogeneity test with $n_1 = n_2 = n_3 = n \in \{ 200, 300, 400, 500, 600\}$. For testing mutual independence, we compare our proposed test to the total multivariance (TM) method \citep{bjorn2019distancemultivariance}.
The significance level is fixed at 0.05.
We use 399 permutation replications to compute $p$-values.
500 Monte Carlo runs are performed to estimate the power. The results are presented in
Figures~\ref{fig:finite_kappa_change}. We also conduct experiments where the sample size $n$ increases but $\kappa$ is fixed, whose results are deferred to supplementary materials. 

From Figure~\ref{fig:finite_kappa_change}A, the power of the MCVM and ED increases to 1 as the gap between distributions enlarges; their power also approaches the significance level as the gap closes. From the upper panel of Figure~\ref{fig:finite_kappa_change}A, when the distributions have a Fr{\'e}chet mean difference, ED outperforms the others, followed by MCVM and FVA that have competitive performance. On the other hand, MCVM is superior in detecting Fr{\'e}chet variance difference. These observations coincide with the finding that the MDF-based method is good at detecting scale differences \citep{kim2018robust}. Notice that the FVA also has a remarkable power at detecting Fr{\'e}chet variance difference for the SPD data, which coincides with the report of \citet{muller2019frechet}. Unfortunately, this fact relies on the well-estimation for the Fr{\'e}chet mean, which may be difficult or even impossible for heavy-tailed data, as we can see in the left-bottom panel in Figure~\ref{fig:finite_kappa_change}A. Moreover, the (approximate) violation of the uniqueness assumption of the Fr{\'e}chet mean hinders the power of the FVA increases. This fact can be witnessed in the middle-bottom panel of Figure~\ref{fig:finite_kappa_change}A --- the power of FVA cannot improve to 1.0 when $\kappa$ exceeds 1.0, where $Y$ approaches a mixture of von-Miser distributions with directions $(0, 1)$ and $(0, -1)$ whose Fr{\'e}chet mean does not exist. 

Figure~\ref{fig:finite_kappa_change}B displays the results of the mutual independence test. From Figure~\ref{fig:finite_kappa_change}B, the power functions of the two tests monotonously increase to 1 as the dependence strength $\kappa$ increases.
Notably, when random objects are mutually independent ($\kappa = 0$), the empirical power of the two tests is around the nominal significance level.
Moreover, TM is better than MA when random objects depend on Fr{\'e}chet mean, but the MA test still enjoys a competitive power. On the other hand, MA is more powerful than TM when random objects have a dependence on variance, which is even more visibly in spherical and SPD data. Lastly, it is noteworthy that the advantages of the MDF-based tests persist in the complex scenarios presented in the final section of the Supplementary Materials. This underscores the promise of our proposed tests for real-world data that may possess complexity. 



\section{Real Data Analysis}\label{sec:real-data}
\subsection{Alzheimer's disease neuroimaging initiative data}
The Alzheimer's Disease Neuroimaging Initiative (ADNI) is a multisite study that aims to
improve the prevention and treatment of Alzheimer's disease (AD). AD is
a neurodegenerative disease, resulting in the decline of some cognitive impairments that cause behavioral deficits.
Data including magnetic resonance images, demographic variables, genetic markers, and AD assessment scale cognitive score (ADASCS) were collected to study AD and the human brain.
In this study, we focus on a critical brain region: the hippocampus, which is typically firstly damaged by AD,
leading to the first clinical manifestations in the form of episodic memory deficits \citep{weiner2013adni}.
By applying our method for the data preprocessed by \citet{kong2018flcrm},
we are interested in factors that affect the hippocampus.

\begin{figure}[t]
	\begin{subfigure}[t]{0.3\textwidth}
		\hspace{-16pt}
		\includegraphics[scale=0.88]{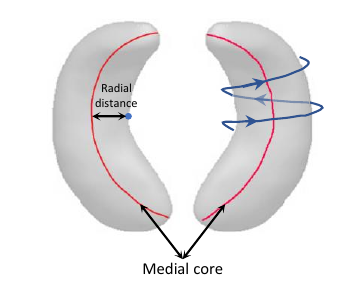}
		\caption{}\label{subfig:radial-distance}
	\end{subfigure}
	\begin{subfigure}[t]{0.7\textwidth}
		\hspace{-16pt}
		\includegraphics[scale=0.5]{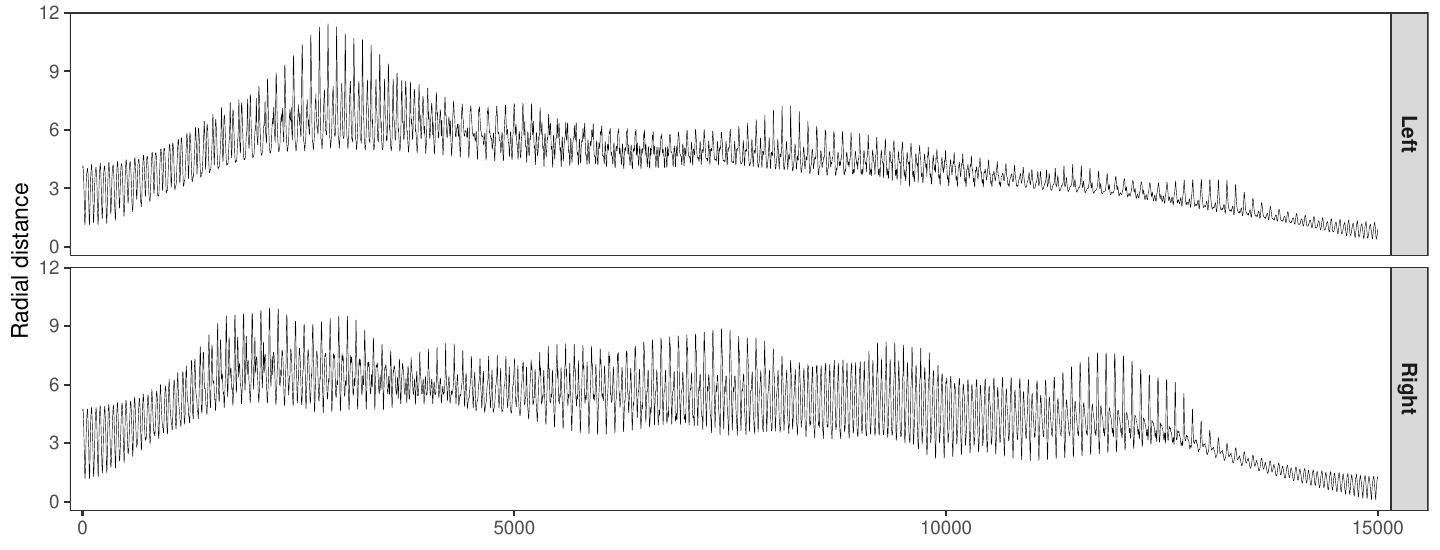}
		\caption{}
	\end{subfigure}\label{subfig:functional-curve}
	\caption{(a) Visualization for medial core, radial distance,
	and the order of 15,000 landmarks on each hippocampus surface.
	(b) One subject's functional curve of radial distance on the left and right hippocampus surfaces.
	The $x$-axis corresponds to 15,000 landmarks that whirlingly surround
	the hippocampus.
	}
	\label{fig:real-radial-distance}
\end{figure}

The preprocessed data contain the left and right hippocampus of 373 individuals,
each of which is characterized by 15,000 radial distances on the left and right hippocampus surfaces,
where the radial distance is defined as the Euclidean distance
between the corresponding vertex on the surface and the medial core of the hippocampus (see Figure~\ref{fig:real-radial-distance}(a)).
From a functional curve example exhibited in Figure~\ref{fig:real-radial-distance}(b),
we see that the functional curve has an obvious fluctuation and periodicity.
And thus, to capture the main variation of functional curves, like \citet{kong2018flcrm},
we apply smoothing and functional principal component analysis \citep{ramsay1997functional} on the functional curves of left and right surfaces, respectively.
We find that the top nine functional principal components for each hippocampus can explain 99\% of the total variance.

We consider gender, age, handedness, marital status, education length, retirement,
Apolipoprotein E (APOE) 3-allele haplotype (i.e., the $\epsilon$2, $\epsilon$3, and $\epsilon$4 variants), and
the ADASCS. The ADASCS is quantitatively evaluated behavioral deficits caused by AD,
and the higher the ADASCS is, the more severe the deficits are.

The metric space for left/right hippocampus is $(L^2, d_{h})$,
where $L^{2}$ is the collection of square integral functions and $d_{h}(u, v) = \| u - v \|_{2}$ for $u, v \in L^2$.
To jointly consider left and right hippocampi, we set
their product metric space as $(L^2 \times L^2, d)$,
where $d(\mathbf{u}, \mathbf{v}) = \| ( d_{h}(u_1, v_1), d_{h}(u_2, v_2) ) \|_{2}$
for $\mathbf{u} = (u_1, u_2), \mathbf{v} = (v_1, v_2) \in L^2 \times L^2$.

To answer the above question, we use the MA and TM tests to evaluate whether demographic and genetic factors affect the hippocampus.
The results are displayed in Table~\ref{tab:hippocampus-testing}.
As seen from Table~\ref{tab:hippocampus-testing}, both MA and TM detect associations of the hippocampus with age and ADASCS.
The increase in age accumulates the abnormal deposition of $\beta$ amyloid fibrils, which starts the neural damage with hippocampus atrophy. Furthermore, the atrophy of the hippocampus causes behavior deficits \citep{hypothetical2010jack}, and the ADASCS are expected to be related to the hippocampus.

\begin{figure}[t]
	\begin{center}
		\includegraphics[scale=0.6]{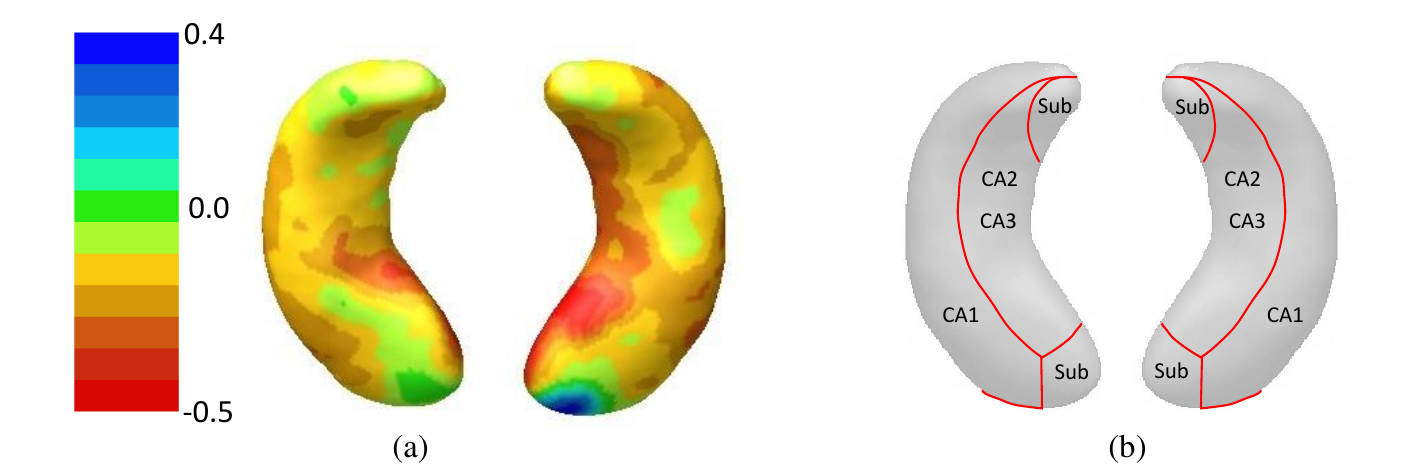}
	\end{center}
	\vspace{-20pt}
	\caption{
		(a) Visualization for mean radial distance difference between APOE$\epsilon$4 and
		non-APOE$\epsilon$4 carriers on hippocampi surfaces.
		The color bar depicts the mean radial distance difference.
		(b) Hippocampal subfields. CA and Sub are abbreviations for cornu amonis and subiculum, respectively.
	}\label{fig:real-radial-distance-manifold}
\end{figure}

Notably, the MA test also detects APOE$\epsilon 4$, but TM does not.
This difference is important because APOE$\epsilon 4$ is a well-known major genetic risk factor for AD
and has repeatedly been reported to affect the hippocampus \citep{hypothetical2010jack}.
Figure~\ref{fig:real-radial-distance-manifold} displays the difference in the mean functional curves between the APOE$\epsilon4$ carriers and non-carriers and indicates that APOE$\epsilon4$ shrinks the hippocampus.
Moreover, Figure~\ref{fig:real-radial-distance-manifold} suggests that the atrophy caused by APOE$\epsilon4$ is more severe in the right hippocampus,
and cornu amonis 2~(CA2) and CA3 are heavily shrunk by APOE$\epsilon4$ following by CA1 and subiculum.
\citet{reduced2012odwyer} also found this phenomenon by studying left and right hippocampi volumes.
\begin{table}[t]\scriptsize
	\caption{
		The $p$-values (adjusted $p$-values under Holm's correction) of independence tests for the ADNI dataset.
		MS and EL are abbreviations of marital status and educational length.
		The $p$-values under 0.05 after Holm's correction are bolded.
	}
	\begin{center}
	\vspace*{-15pt}
	\begin{tabular}{c|ccccc}
		\toprule
		 Test & Gender & Handedness & MS & EL & Retirement  \\
		\midrule
		\multirow{2}{*}{}
		TM            & 0.012 (0.096) & 0.760 (1.000) & 0.628 (1.000) & 0.020 (0.140) & 0.318 (1.000) \\
		MA (Permute)  & 0.038 (0.259) & 0.589 (1.000) & 0.830 (1.000) & 0.079 (0.395) & 0.507 (1.000) \\
		MA (Spectrum) & 0.046 (0.295) & 0.546 (1.000) & 0.811 (1.000) & 0.077 (0.386) & 0.499 (1.000) \\
		\midrule
 		Test & Age & APOE$\epsilon 2$ & APOE$\epsilon 3$ & APOE$\epsilon 4$  & ADASCS \\
		\midrule
		\multirow{2}{*}{}
            TM            & \textbf{0.001 (0.010)}  & 0.871 (1.000) & 0.228 (1.000) & 0.023 (0.140)          & \textbf{0.001 (0.010) } \\ 
            MA (Permute)  & \textbf{0.001 (0.010)}  & 0.226 (0.904) & 0.037 (0.259) & \textbf{0.004 (0.032)} & \textbf{0.001 (0.010) } \\ 
            MA (Spectrum) & \textbf{0.0003 (0.001)} & 0.229 (0.918) & 0.042 (0.295) & \textbf{0.002 (0.018)} & \textbf{0.0002 (0.002) } \\ 
		\bottomrule
	\end{tabular}\label{tab:hippocampus-testing}
	\vspace*{-10pt}
	\end{center}
\end{table}     

\subsection{ADHD-200 dataset}\label{sec:real-data-adhd200}
The correlated spontaneous fluctuations manifest different patterns in different brain regions
during sleep or under anesthesia. This phenomenon is called functional connectome (FC), where a node represents a part of the brain, and an edge represents the direct correlation between two nodes. It is interesting to evaluate the relationship between FC phenotypes and other variables. For example, we investigate the mutual dependence among FC, gender, and handedness with ADHD-200 dataset \citep{adhd2017pierre}.

The ADHD-200 dataset includes 162 individuals, with 63 males and 99 females. In this dataset, handedness is measured as a continuous score --- a person with large handedness implies that he/she is dextromanuality. 
For each subject, the ADHD-200 dataset provides a preprocessed resting-state functional magnetic resonance imaging (rfMRI) that repeatedly records the blood-oxygen levels on 111 disjoint regions in the brain. 
As suggested by \citet{smith2013functional}, we can reconstruct the functional connectome by computing a $111\times 111$ partial correlation matrix of the disjoint regions. Then, the Cholesky distance is chosen to measure the difference among SPDs due to its favorable performance in the simulation studies. 

We applied the MA-based and TM tests to this dataset to answer our question. The test results are displayed in Table~\ref{tab:adhd200}. As can be seen from the first column of Table~\ref{tab:adhd200}, only the MA-based tests reject the null hypothesis of the mutual independence among FC, gender, and handedness, while the TM test does not. In addition, we divide the three factors into two groups and study the dependence of the two groups. The results are presented in the third to fifth columns in Table~\ref{tab:adhd200}. The third column shows that gender and handedness together may affect the FC by our MA-based methods.  While the null hypothesis of the independence between (gender, FC) and handedness is not rejected in the fourth column, the null hypothesis of the independence between (handedness, FC) and gender is hard to judge in the fifth column. We further investigate the pairwise dependence among three factors. From the sixth column in Table~\ref{tab:adhd200}, we see that only the MA-based tests reject the hypothesis of the independence between handedness and FC. A recent study found that the FCs of left- and right-handed individuals extend across every brain region \citep{tejavibulya2022large}. They discovered that connections between and within the cerebellum have distinct connectivity patterns. 
Besides, from the seventh to eighth columns in Table~\ref{tab:adhd200}, both the MA-based and TM tests suggest insufficient evidence to assert gender influences handedness or FC. 
In conclusion, our MA-based method for mutual independence is very useful for exploring the high-order dependence among variables, including non-Euclidean ones.


\begin{table}[t]\scriptsize
	\caption{
		The $p$-values (adjusted $p$-values under Holm's correction) of (mutual) independence tests for the ADHD-200 dataset. The $p$-values under 0.05 after Holm's correction are bolded. 
        G, H, and FC are abbreviations of gender, handedness, and functional connectome, respectively. 
        MA$_1$: spectrum-based test; MA$_2$: permutation-based test. Note that the results are rounded to three digits.
	}
	\begin{center}
	\vspace*{-15pt}
        \begin{tabular}{c|c|ccc|ccc}
        	\toprule
        	Test & Joint & (G, H)-FC & (G, FC)-H & (H, FC)-G & FC-H & FC-G & H-G   \\
        	\midrule
        	TM               & 0.206 & 0.018 (0.054) & 0.321 (0.321) & 0.037 (0.074) & 0.020 (0.060) & 0.057 (0.114) & 0.755 (0.755) \\
        	MA$_1$ & \textbf{0.017} & \textbf{0.006 (0.018)} & 0.642 (0.642) & \textbf{0.017 (0.035)} & \textbf{0.005 (0.016)} & 0.112 (0.224) & 0.989 (0.989) \\ 
        	MA$_2$ & \textbf{0.022} & \textbf{0.005 (0.015)} & 0.701 (0.701) & \textbf{0.017 (0.034)} & \textbf{0.008 (0.024)} & 0.139 (0.278) & 0.876 (0.876) \\
        	\bottomrule
        \end{tabular}\label{tab:adhd200}
	\vspace*{-10pt}
	\end{center}
\end{table}

\section{Conclusion and Discussion}\label{sec:summary}
To address the need for non-Euclidean data analysis, we characterize Borel probability measures in metric space. This is analogous to how the DF represents Borel probability measures in Euclidean space. We propose a metric distribution function. To a large extent, the MDF retains the DF's desirable properties, including the 1-1 correspondence theorem, under certain mild conditions. Moreover, the EMDF provides a simple and reliable method for estimating the MDF, which coincides with the weak mode of calculating the probability measure in \cite{vapnik2013nature}. The outer-directed loop in Figure~\ref{inference-framework} is closed thanks to the EMDF's Glivenko-Cantelli property and the Donsker property; this paves the way for a new paradigm in statistical inference in metric spaces. As a result, we can perform nonparametric statistical inference for non-Euclidean data with the MDF, similar to that for Euclidean data with the DF.

We provide extensive empirical results for using the MDF in the homogeneity test and the (mutual) independence test for non-Euclidean data, including SPD matrices, shapes, and smooth functions, to highlight the great potential of MDF in practice. Using the Glivenko-Cantelli property and the Donsker property of the EMDF, we obtain estimators and their consistency for estimations and test procedures that are described in Section~\ref{sec:statistical-method}. Our simulation experiments demonstrate that the tests based on the MDF have good finite sample performance, are robust to various data settings, and are free of tuning parameters. Therefore, the applicability of MDF is expected to be broad and simple. We also re-analyzed the ADNI data set and found evidence that the APOE$\epsilon$4 affected the hippocampus.

The MDF and the DF are connected. Denote $F(\mathbf{u})=P(X\leq \mathbf{u})$ as the DF associated with probability measure $\mu$ defined in $\mathbb{R}$. Given $\mathbf{u}, \mathbf{v}\in \mathbb{R}$ such that $\mathbf{u}<\mathbf{v}$, we have  
     \begin{align*}
        F_\mu^M(\mathbf{u},\mathbf{v}) = P(|X-\mathbf{u}|\leq |\mathbf{v}-\mathbf{u}|)= P(X  \leq \mathbf{v})-P(X  < 2\mathbf{u}-\mathbf{v}).
     \end{align*}
Let $\mathbf{u}\rightarrow-\infty,$ then we have $\lim\limits_{\mathbf{u}\rightarrow-\infty}F_\mu^M(\mathbf{u},\mathbf{v})=F(\mathbf{u})$. Despite their connections, the MDF and the DF are not the same. 
The MDF describes the random object by the distribution of distances at different locations.
Thus, the viewpoint of MDF is flexible and relative since the location variable $\mathbf{u}$ can be different.
Such characteristics allow the MDF to grasp the information of probability measures in metric spaces.

Notably, the MDF is also related to another useful concept in nonparametric statistics --- statistical depth. The statistical depth methods use a unique center, the so-called deepest point, to characterize the distribution. These methods can properly characterize distributions in Euclidean spaces, and recent progress enables the characteristics of the depth in metric space \citep{daiTukeyDepthObject2021, liu2022quantiles}. We note that the method proposed by \citet{liu2022quantiles} is derived from the MDF, providing another example that the MDF establishes a unified framework for analyzing complex data objects.
             


\bigskip
\begin{center}
{\large\bf Supplementary materials}
\end{center}

\begin{description}
\item[Supplementary document:] 
contains technical proof, some properties of EMDF, a discussion about the spectrum-based tests, and additional simulation results. (.pdf file)



\end{description}


\bibliographystyle{Chicago}
\bibliography{reference}

\begin{thebibliography}{}

\bibitem[\protect\citeauthoryear{Bellec, Chu, Chouinard-Decorte, Benhajali,
  Margulies, and Craddock}{Bellec et~al.}{2017}]{adhd2017pierre}
Bellec, P., C.~Chu, F.~Chouinard-Decorte, Y.~Benhajali, D.~S. Margulies, and
  R.~C. Craddock (2017).
\newblock The neuro bureau {ADHD}-200 preprocessed repository.
\newblock {\em NeuroImage\/}~{\em 144}, 275 -- 286.
\newblock Data Sharing Part II.

\bibitem[\protect\citeauthoryear{Böttcher, Keller-Ressel, and
  Schilling}{Böttcher et~al.}{2019}]{bjorn2019distancemultivariance}
Böttcher, B., M.~Keller-Ressel, and R.~L. Schilling (2019).
\newblock Distance multivariance: New dependence measures for random vectors.
\newblock {\em Annals of Statistics\/}~{\em 47\/}(5), 2757 -- 2789.

\bibitem[\protect\citeauthoryear{Castillo, Schmidt-Hieber, Van~der Vaart,
  et~al.}{Castillo et~al.}{2015}]{Castillo2015Bayesian}
Castillo, I., J.~Schmidt-Hieber, A.~Van~der Vaart, et~al. (2015).
\newblock Bayesian linear regression with sparse priors.
\newblock {\em Annals of Statistics\/}~{\em 43\/}(5), 1986--2018.

\bibitem[\protect\citeauthoryear{Cornea, Zhu, Kim, Ibrahim, and
  Initiative}{Cornea et~al.}{2017}]{cornea2017regression}
Cornea, E., H.~Zhu, P.~Kim, J.~G. Ibrahim, and A.~D.~N. Initiative (2017).
\newblock Regression models on riemannian symmetric spaces.
\newblock {\em Journal of the Royal Statistical Society: Series B (Statistical
  Methodology)\/}~{\em 79\/}(2), 463--482.

\bibitem[\protect\citeauthoryear{Dai and {Lopez-Pintado}}{Dai and
  {Lopez-Pintado}}{2022}]{daiTukeyDepthObject2021}
Dai, X. and S.~{Lopez-Pintado} (2022).
\newblock Tukey's depth for object data.
\newblock {\em Journal of the American Statistical Association\/}, 1--13.

\bibitem[\protect\citeauthoryear{Darling}{Darling}{1957}]{darling1957the}
Darling, D.~A. (1957).
\newblock The {K}olmogorov-{S}mirnov, {C}ram{\'e}r-von {M}ises tests.
\newblock {\em The Annals of Mathematical Statistics\/}~{\em 28\/}(4),
  823--838.

\bibitem[\protect\citeauthoryear{Davies}{Davies}{1971}]{davies1971measures}
Davies, R.~O. (1971).
\newblock Measures not approximable or not specifiable by means of balls.
\newblock {\em Mathematik\/}~{\em 18\/}(02), 157--160.

\bibitem[\protect\citeauthoryear{Do~Carmo and Flaherty~Francis}{Do~Carmo and
  Flaherty~Francis}{1992}]{do1992riemannian}
Do~Carmo, M.~P. and J.~Flaherty~Francis (1992).
\newblock {\em Riemannian geometry}, Volume~6.
\newblock Springer.

\bibitem[\protect\citeauthoryear{Driemel, Nusser, Phillips, and
  Psarros}{Driemel et~al.}{2021}]{driemel2019vc}
Driemel, A., A.~Nusser, J.~M. Phillips, and I.~Psarros (2021).
\newblock The {VC} dimension of metric balls under {F}r{\'e}chet and
  {H}ausdorff distances.
\newblock {\em Discrete \& Computational Geometry\/}~{\em 66\/}(4), 1351--1381.

\bibitem[\protect\citeauthoryear{Dryden, Koloydenko, and Zhou}{Dryden
  et~al.}{2009}]{dryden2009noneuclidean}
Dryden, I.~L., A.~Koloydenko, and D.~Zhou (2009, 09).
\newblock Non-euclidean statistics for covariance matrices, with applications
  to diffusion tensor imaging.
\newblock {\em Annals of Applied Statistics\/}~{\em 3\/}(3), 1102--1123.

\bibitem[\protect\citeauthoryear{Dubey and Muller}{Dubey and
  Muller}{2019}]{muller2019frechet}
Dubey, P. and H.-G. Muller (2019, 10).
\newblock Fr{\'e}chet analysis of variance for random objects.
\newblock {\em Biometrika\/}~{\em 106\/}(4), 803--821.

\bibitem[\protect\citeauthoryear{Efron}{Efron}{1979}]{10.1214/aos/1176344552}
Efron, B. (1979).
\newblock Bootstrap methods: Another look at the jackknife.
\newblock {\em Annals of Statistics\/}~{\em 7\/}(1), 1--26.

\bibitem[\protect\citeauthoryear{Federer}{Federer}{2014}]{federer2014geometric}
Federer, H. (2014).
\newblock {\em Geometric measure theory}.
\newblock Berlin, Heidelberg: Springer.

\bibitem[\protect\citeauthoryear{Halmos}{Halmos}{1956}]{halmos1956measure}
Halmos, P.~R. (1956).
\newblock {\em Measure theory}.
\newblock New York: D. Van Nostrand Company, Inc.

\bibitem[\protect\citeauthoryear{Hoeffding}{Hoeffding}{1948}]{hoeffding1948independence}
Hoeffding, W. (1948, 12).
\newblock A non-parametric test of independence.
\newblock {\em The Annals of Mathematical Statistics\/}~{\em 19\/}(4),
  546--557.

\bibitem[\protect\citeauthoryear{Hong, Kwitt, Singh, Vasconcelos, and
  Niethammer}{Hong et~al.}{2016}]{HongMarc2016}
Hong, Y., R.~Kwitt, N.~Singh, N.~Vasconcelos, and M.~Niethammer (2016).
\newblock Parametric regression on the grassmannian.
\newblock {\em IEEE Transactions on Pattern Analysis and Machine
  Intelligence\/}~{\em 38}, 2284--2297.

\bibitem[\protect\citeauthoryear{Jack, Knopman, Jagust, Shaw, Aisen, Weiner,
  Petersen, and Trojanowski}{Jack et~al.}{2010}]{hypothetical2010jack}
Jack, C.~R., D.~S. Knopman, W.~J. Jagust, L.~M. Shaw, P.~S. Aisen, M.~W.
  Weiner, R.~C. Petersen, and J.~Q. Trojanowski (2010).
\newblock Hypothetical model of dynamic biomarkers of the alzheimer's
  pathological cascade.
\newblock {\em The Lancet Neurology\/}~{\em 9\/}(1), 119--128.

\bibitem[\protect\citeauthoryear{Kim, Balakrishnan, and Wasserman}{Kim
  et~al.}{2020}]{kim2018robust}
Kim, I., S.~Balakrishnan, and L.~Wasserman (2020).
\newblock {Robust multivariate nonparametric tests via projection averaging}.
\newblock {\em Annals of Statistics\/}~{\em 48\/}(6), 3417 -- 3441.

\bibitem[\protect\citeauthoryear{Kong, Ibrahim, Lee, and Zhu}{Kong
  et~al.}{2018}]{kong2018flcrm}
Kong, D., J.~G. Ibrahim, E.~Lee, and H.~Zhu (2018).
\newblock {FLCRM}: Functional linear cox regression model.
\newblock {\em Biometrics\/}~{\em 74\/}(1), 109--117.

\bibitem[\protect\citeauthoryear{Lin and M{\"u}ller}{Lin and
  M{\"u}ller}{2021}]{lin2019total}
Lin, Z. and H.-G. M{\"u}ller (2021).
\newblock Total variation regularized fr{\'e}chet regression for metric-space
  valued data.
\newblock {\em Annals of Statistics\/}~{\em 49\/}(6), 3510 -- 3533.

\bibitem[\protect\citeauthoryear{Liu, Wang, and Zhu}{Liu
  et~al.}{2022}]{liu2022quantiles}
Liu, H., X.~Wang, and J.~Zhu (2022).
\newblock Quantiles, ranks and signs in metric spaces.
\newblock {\em arXiv preprint arXiv:2209.04090\/}.

\bibitem[\protect\citeauthoryear{O'Dwyer, Lamberton, Matura, Tanner, Scheibe,
  Miller, Rujescu, Prvulovic, and Hampel}{O'Dwyer
  et~al.}{2012}]{reduced2012odwyer}
O'Dwyer, L., F.~Lamberton, S.~Matura, C.~Tanner, M.~Scheibe, J.~Miller,
  D.~Rujescu, D.~Prvulovic, and H.~Hampel (2012, 11).
\newblock Reduced hippocampal volume in healthy young apoe4 carriers: An {MRI}
  study.
\newblock {\em PLOS ONE\/}~{\em 7\/}(11), 1--10.

\bibitem[\protect\citeauthoryear{O'Hara, Sillanp{\"a}{\"a}, et~al.}{O'Hara
  et~al.}{2009}]{o2009review}
O'Hara, R.~B., M.~J. Sillanp{\"a}{\"a}, et~al. (2009).
\newblock A review of bayesian variable selection methods: What, how and which.
\newblock {\em Bayesian Analysis\/}~{\em 4\/}(1), 85--117.

\bibitem[\protect\citeauthoryear{Pan, Tian, Wang, and Zhang}{Pan
  et~al.}{2018}]{pan2018ball}
Pan, W., Y.~Tian, X.~Wang, and H.~Zhang (2018, 06).
\newblock Ball divergence: Nonparametric two sample test.
\newblock {\em Annals of Statistics\/}~{\em 46\/}(3), 1109--1137.

\bibitem[\protect\citeauthoryear{{Pan}, {Wang}, {Zhang}, {Zhu}, and
  {Zhu}}{{Pan} et~al.}{2020}]{pan2020ball}
{Pan}, W., X.~{Wang}, H.~{Zhang}, H.~{Zhu}, and J.~{Zhu} (2020).
\newblock Ball covariance: A generic measure of dependence in banach space.
\newblock {\em Journal of the American Statistical Association\/}~{\em
  115\/}(529), 307--317.

\bibitem[\protect\citeauthoryear{Petersen, Liu, and Divani}{Petersen
  et~al.}{2021}]{petersen2021wasserstein}
Petersen, A., X.~Liu, and A.~A. Divani (2021).
\newblock Wasserstein $f$-tests and confidence bands for the fr{\'e}chet
  regression of density response curves.
\newblock {\em Annals of Statistics\/}~{\em 49\/}(1), 590--611.

\bibitem[\protect\citeauthoryear{Petersen}{Petersen}{2006}]{petersen2006riemannian}
Petersen, P. (2006).
\newblock {\em Riemannian geometry}, Volume 171.
\newblock Springer.

\bibitem[\protect\citeauthoryear{Ramsay and Silverman}{Ramsay and
  Silverman}{1997}]{ramsay1997functional}
Ramsay, J. and B.~W. Silverman (1997).
\newblock {\em Functional Data Analysis}.
\newblock New York: Springer.

\bibitem[\protect\citeauthoryear{Scealy and Wood}{Scealy and
  Wood}{2019}]{wood2019scaled}
Scealy, J.~L. and A.~T.~A. Wood (2019).
\newblock Scaled von mises–fisher distributions and regression models for
  paleomagnetic directional data.
\newblock {\em Journal of the American Statistical Association\/}~{\em
  114\/}(528), 1547--1560.

\bibitem[\protect\citeauthoryear{Smith, Vidaurre, Beckmann, Glasser, Jenkinson,
  Miller, Nichols, Robinson, Salimi-Khorshidi, Woolrich, Barch, U{\u g}urbil,
  and Essen}{Smith et~al.}{2013}]{smith2013functional}
Smith, S.~M., D.~Vidaurre, C.~F. Beckmann, M.~F. Glasser, M.~Jenkinson, K.~L.
  Miller, T.~E. Nichols, E.~C. Robinson, G.~Salimi-Khorshidi, M.~W. Woolrich,
  D.~M. Barch, K.~U{\u g}urbil, and D.~C.~V. Essen (2013).
\newblock Functional connectomics from resting-state {fMRI}.
\newblock {\em Trends in Cognitive Sciences\/}~{\em 17\/}(12), 666 -- 682.
\newblock Special Issue: The Connectome.

\bibitem[\protect\citeauthoryear{Sz{\'e}kely and Rizzo}{Sz{\'e}kely and
  Rizzo}{2004}]{szekely2004testing}
Sz{\'e}kely, G.~J. and M.~L. Rizzo (2004).
\newblock Testing for equal distributions in high dimension.
\newblock {\em InterStat\/}~{\em 5\/}(16.10), 1249--1272.

\bibitem[\protect\citeauthoryear{Sz{\'e}kely, Rizzo, and Bakirov}{Sz{\'e}kely
  et~al.}{2007}]{szekely2007measuring}
Sz{\'e}kely, G.~J., M.~L. Rizzo, and N.~K. Bakirov (2007).
\newblock Measuring and testing dependence by correlation of distances.
\newblock {\em Annals of Statistics\/}~{\em 35\/}(6), 2769--2794.

\bibitem[\protect\citeauthoryear{Tejavibulya, Peterson, Greene, Gao, Rolison,
  Noble, and Scheinost}{Tejavibulya et~al.}{2022}]{tejavibulya2022large}
Tejavibulya, L., H.~Peterson, A.~Greene, S.~Gao, M.~Rolison, S.~Noble, and
  D.~Scheinost (2022).
\newblock Large-scale differences in functional organization of left-and
  right-handed individuals using whole-brain, data-driven analysis of
  connectivity.
\newblock {\em NeuroImage\/}~{\em 252}, 119040.

\bibitem[\protect\citeauthoryear{Vapnik}{Vapnik}{2010}]{vapnik2013nature}
Vapnik, V. (2010).
\newblock {\em The Nature of Statistical Learning Theory}.
\newblock New York: Springer.

\bibitem[\protect\citeauthoryear{Wainwright}{Wainwright}{2019}]{wainwright_2019}
Wainwright, M.~J. (2019).
\newblock {\em High-Dimensional Statistics: A Non-Asymptotic Viewpoint}.
\newblock Cambridge: Cambridge University Press.

\bibitem[\protect\citeauthoryear{Weiner, Veitch, Aisen, Beckett, Cairns, Green,
  Harvey, Jack, Jagust, Liu, Morris, Petersen, Saykin, Schmidt, Shaw, Shen,
  Siuciak, Soares, Toga, and Trojanowski}{Weiner et~al.}{2013}]{weiner2013adni}
Weiner, M.~W., D.~P. Veitch, P.~S. Aisen, L.~A. Beckett, N.~J. Cairns, R.~C.
  Green, D.~Harvey, C.~R. Jack, W.~Jagust, E.~Liu, J.~C. Morris, R.~C.
  Petersen, A.~J. Saykin, M.~E. Schmidt, L.~Shaw, L.~Shen, J.~A. Siuciak,
  H.~Soares, A.~W. Toga, and J.~Q. Trojanowski (2013).
\newblock The alzheimer's disease neuroimaging initiative: A review of papers
  published since its inception.
\newblock {\em Alzheimer's \& Dementia\/}~{\em 9\/}(5), e111--e194.

\bibitem[\protect\citeauthoryear{Wellner et~al.}{Wellner
  et~al.}{2013}]{wellner2013weak}
Wellner, J. et~al. (2013).
\newblock {\em Weak convergence and empirical processes: with applications to
  statistics}.
\newblock Springer Science \& Business Media.

\end{thebibliography}
\end{document}